\documentclass[3p,12pt]{elsarticle}

\usepackage{graphicx}
\usepackage[tight]{subfigure}
\usepackage{amssymb}
\usepackage{amsmath}
\usepackage{multirow}

\usepackage[utf8]{inputenc}

\journal{}

\begin{document}

\begin{frontmatter}

%% Title, authors and addresses

%% use the tnoteref command within \title for footnotes;
%% use the tnotetext command for the associated footnote;
%% use the fnref command within \author or \address for footnotes;
%% use the fntext command for the associated footnote;
%% use the corref command within \author for corresponding author footnotes;
%% use the cortext command for the associated footnote;
%% use the ead command for the email address,
%% and the form \ead[url] for the home page:
%%
%% \title{Title\tnoteref{label1}}
%% \tnotetext[label1]{}
%% \author{Name\corref{cor1}\fnref{label2}}
%% \ead{email address}
%% \ead[url]{home page}
%% \fntext[label2]{}
%% \cortext[cor1]{}
%% \address{Address\fnref{label3}}
%% \fntext[label3]{}

\title{SPH Entropy Errors and the Pressure Blip}
%% use optional labels to link authors explicitly to addresses:
%% \author[label1,label2]{<author name>}
%% \address[label1]{<address>}
%% \address[label2]{<address>}

\author[IITB]{Kunal Puri\corref{cor1}}
\ead{kunal.r.puri@gmail.com} 

\author[IITB]{Prabhu Ramachandran}
\address[IITB]{Department of Aerospace Engineering, Indian Institute of
  Technology Bombay, Powai, Mumbai 400076}

\cortext[cor1]{Corresponding author}

%\address{Department of Aerospace Engineering, Indian Institute of
%  Technology Bombay, Powai, Mumbai 400076}

%\author{}
%\address{}

\begin{abstract}
The spurious pressure jump at a contact discontinuity, in SPH
simulations of the compressible Euler equations is investigated. From
the spatiotemporal behaviour of the error, the SPH pressure jump is
likened to entropy errors observed for artificial viscosity based
finite difference/volume schemes. The error is observed to be
generated at start-up and dissipation is the only recourse to mitigate
it's effect.\newline We show that similar errors are generated for the
Lagrangian plus remap version of the Piecewise Parabolic Method (PPM)
finite volume code (PPMLR). Through a comparison with the direct
Eulerian version of the PPM code (PPMDE), we argue that a lack of
diffusion across the material wave (contact discontinuity) is
responsible for the error in PPMLR. We verify this hypothesis by
constructing a more dissipative version of the remap code using a
piecewise constant reconstruction. As an application to SPH, we
propose a hybrid GSPH scheme that adds the requisite dissipation by
utilizing a more dissipative Riemann solver for the energy
equation. The proposed modification to the GSPH scheme, and it's
improved treatment of the anomaly is verified for flows with strong
shocks in one and two dimensions. \newline The result that dissipation
must act across the density and energy equations provides a consistent
explanation for many of the hitherto proposed ``cures'' or ``fixes''
for the problem.
\end{abstract}

\begin{keyword}
%% keywords here, in the form: keyword \sep keyword
{SPH}, {GSPH}, {Pressure wiggling}, {Entropy errors}

%% MSC codes here, in the form: \MSC code \sep code
%% or \MSC[2008] code \sep code (2000 is the default)

\end{keyword}

\end{frontmatter}

%%
%% Start line numbering here if you want
%%
% \linenumbers

%% main text
%% The Appendices part is started with the command \appendix;
%% appendix sections are then done as normal sections
%% \appendix

%% \section{}
%% \label{}

\section{Introduction}
\label{sec:entropy-errors}
SPH solutions to the compressible Euler equations are characterized by
an anomalous ``blip'' or kink in the pressure at the contact
discontinuity. The density profile is accurate which means the
internal energy shows a corresponding heating/cooling to mirror the
pressure jump. The error, once introduced, neither grows nor
attenuates without dissipation and is simply advected with the
particles at the local material velocity. Monaghan and
Gingold~\cite{monaghan-gingold-shock} were the first to observe this
behaviour when they applied SPH to simulate shock-tube problems.
Their observations lead them to ascribe the phenomenon to general
``starting'' errors when discontinuous initial profiles are
used. Presumably, SPH struggles with the discontinuous thermal
energy. Resolving this behaviour has been the focus of numerous
researchers over the last thirty years as this is manifestly a grave
drawback of the method. Despite this error, SPH has been found to be
useful within the astrophysics community, it's application often
preceded by ``code-comparisons'' with existing Eulerian
techniques. One of the early comparisons was undertaken by Davies et
al.~\cite{davies93}, who compared SPH simulations of stellar
collisions with the Piecewise Parabolic Method (PPM). They suggest
that the advantages of each approach are mutually exclusive, although
the two approaches were qualitatively similar. Caution is advised in
extending this observation for other calculations in which different
hydrodynamic effects determine the solution. About the same time,
Steinmetz and M\"uller~\cite{steinmetz-muller1993} had also suggested
that SPH and finite difference methods should be looked upon as
complimentary methods to solve hydrodynamic problems. In their seminal
work, Agertz et al.~\cite{sphvgrids} performed a comprehensive
comparison of astrophysical codes (using GADGET~\cite{gadget-code} for
SPH) for the simulation of interacting multi-phase fluids. The
un-physical pressure jump at a density gradient, as produced by SPH in
it's standard formulation was found to render the method incapable of
resolving hydrodynamic instabilities like the Kelvin-Helmholtz (KHI)
or Rayleigh-Taylor instabilities (RTI). A similar comparison was
carried out by Tasker et al.~\cite{taskeretal08} for test problems
with analytical solutions, therefore enabling a more quantitative
comparison. While SPH was found to be generally comparable in it's
accuracy with the Eulerian schemes, a major difference was the
pressure jump at the contact discontinuity, which is absent for
grid-based codes. For the hydrodynamics of multi-phase fluids (more
generally at a density gradient), this spurious pressure jump behaves
like an artificial surface tension force, inhibiting the development
of density driven instabilities like KHI. In another study, Okamoto et
al.~\cite{okamoto03} observed that the erroneous pressure jump can
also result in spurious momentum transfer across shearing flows,
significantly affecting numerical results. These code comparisons
rekindled the need to resolve the spurious pressure at the contact
discontinuity, with the Kelvin-Helmholtz instability (KHI) often used
as a canonical ``mixing'' problem exposing the method's
vulnerability. \newline Among the many tricks for
SPH~\cite{herant-tricks}, arguably the oldest one is a judicious use
of artificial dissipation. Thermal conduction is as old as artificial
viscosity itself with Monaghan~\cite{monaghan-review} and
Brookshaw~\cite{brookshaw94} being early advocates for it's use in
treating ``wall-heating'' errors. It has been used for example, by
Sigalotti et al.~\cite{sigalotti2006-shock,sigalotti2008-shock} and
Rosswog and Price~\cite{rosswog-price07} for strong shock problems in
hydrodynamics and magnetohydrodynamics respectively. Addressing the
mixing problem originally highlighted by Agertz et
al.~\cite{sphvgrids}, Price~\cite{price2008-khi} demonstrated that a
judicious use of thermal conduction enables a suitable description of
density driven instabilities like KHI. The conduction terms are
formulated using the signal-based artificial
viscosity~\cite{monaghan1997-riemann} and are constructed to result in
a diffusion of energy across the contact discontinuity. The use of and
need for similar thermal conduction terms was also suggested by
Wadsley et al.~\cite{Wadsley08}, Garc\'ia-Senz et
al.~\cite{Garcia-Senz01012009} and and Valcke et al.~\cite{valcke10}
for mixing problems in astrophysics. Some authors also suggest that
apart from the use of the thermal conduction, the magnitude of the
pressure jump can be curtailed by relaxing the initial conditions and
by using a modified kernel with an increased sampling. Thermal
conduction is necessary for the long term simulation and to avoid
``oily''~\cite{valcke10} or ``gloopy''~\cite{read10} features in the
solution. Following Price, Valdarnini~\cite{valdarnini12}, Kawata et
al.~\cite{kawata13} and Rosswog~\cite{rosswog09-review} have also
advocated the use of artificial thermal conduction. By using an error
and stability analysis, Read et al.~\cite{read10} showed that the
inability of SPH to adequately resolve mixing was due in part to a
``Local Mixing Instability'' (LMI), whereby, particles are inhibited
to mix on the kernel scale due to entropy conservation, which in turn
results in a pressure discontinuity. The LMI is therefore another term
for the pressure ``blip'' in the context of hydrodynamic mixing. The
LMI was cured by using a modified density estimate, similar to that
employed by Ritchie and Thomas~\cite{RT01}, to ensure a single valued
pressure throughout the flow. The modified density approaches
(\cite{RT01,MarriWhite03,read10}) are designed for a more accurate
density estimation for multi-phase fluids (mixing problems) in
pressure equilibrium. Consequently, they perform poorly for flows with
strong shocks. Indeed, in a recent article, Read and
Hayfield~\cite{read12} discuss a new high-order dissipation switch for
adaptive viscosity in which they forgo the modified density approach
in favour of an artificial heating term as proposed by
Price~\cite{price2008-khi}.\newline

Moving away from adding thermal conduction in a somewhat ad-hoc
manner, Price~\cite{price2012} argues that the assumption of a
differentiable density is the cause of the spurious pressure jump. The
density estimate plays a central role in the variational formulation
of SPH and is used to define an implicit particle volume through the
ratio of particle mass to particle density. Saitoh and
Makino~\cite{saitoh12} took cue from this idea to develop a
density-independent SPH (DISPH) by replacing the mass density by an
equivalent pressure density and it's arbitrary
function. Hopkins~\cite{hopkins12} also considered the idea of
replacing the particle volume, traditionally defined by the mass
density, by an arbitrary smoothed function. A family of equivalent
Lagrangian schemes are derived by different choices of the
function. In particular, the pressure-entropy formulation was shown to
be superior at resolving mixing in the hydrodynamic context. However,
there appear to be problems for shocked flows (due to the
non-isentropic nature of the flow), similar to the modified density
approach occurs for this formulation as well~\cite{sphgal}.\newline

The SPH formulations discussed hitherto were of the variational kind
with the use of explicit dissipation
terms. Inutsuka~\cite{inutsuka-riemann} developed an
artificial-viscosity free scheme that requires the solution of a
Riemann problem between interacting particle pairs. The Riemann solver
introduces the necessary and sufficient dissipation required to
stabilize the scheme. Although the pressure blip is present for these
Godunov SPH (GSPH) schemes, it is less pronounced. The result is a
more suitable description of fluid instabilities like KHI. Indeed, Cha
et al.~\cite{cha-inutsuka-khi} found GSPH to be superior to the
standard SPH for the development of KHI. They argue in favour of the
linear consistency in the GSPH momentum equation and a more accurate
Lagrangian function used in GSPH. Murante et al.~\cite{murante-gsph}
observed similar advantages of GSPH for the simulation of
hydrodynamical instabilities vis-a-vis standard SPH. Another
artificial-viscosity free SPH scheme was proposed by
Lanzafame~\cite{lanzafame09} by considering shock flows as
non-equilibrium events. The equation of state is reformulated
according to a Riemann problem to introduce the necessary
dissipation. Incidentally, these GodunovSPH schemes typically include
an implicit thermal conduction term which is known
(\cite{price2008-khi}) to ameliorate the pressure jump. Indeed, the
authors have shown~\cite{purigsph13} that GSPH with a class of
approximate Riemann solvers is essentially equivalent to the standard
SPH with artificial viscosity and thermal conduction terms.\newline 

It is worth noting that despite the numerous efforts to address this
pressure jump, the error is at best ameliorated, with it's adverse
effects kept to within acceptable limits. Care must be exercised with
the use of thermal conduction so as to avoid excessive smearing and a
resulting loss of accuracy. This is achieved through viscosity
limiters~\cite{cullen-walter,read12,taylor12,price2012} and solution
dependent conductivity coefficients
\cite{sigalotti2008-shock,price2012}. Without a unifying theory
however, the different approaches seem serendipitous and somewhat
ad-hoc. Success of a particular method notwithstanding, a discernible
pattern among all proposed solutions is the introduction of a certain
dissipation into the equations of motion to handle the pressure
jump. The dissipation is introduced either in the form of an explicit
(\cite{springel2010,price2008-khi,price2012}) or implicit (GSPH)
thermal conduction
(\cite{inutsuka-riemann,cha-inutsuka-khi,murante-gsph}), or by more
subtle means via the state equation \cite{lanzafame09}, density
estimate~\cite{read10} or particle
volume~\cite{saitoh12,hopkins12}.\newline That dissipation can be used
to progressively smear the pressure blip once it is created should be
fairly obvious. A more fundamental question that can be asked perhaps,
pertains to the origin of this error. Towards this goal, we search for
similar behaviour in finite difference/volume schemes. These
grid-based schemes have received a great deal of attention and success
within the CFD community and it is therefore helpful to study them. In
particular, if we can relate the SPH errors to those generated with a
suitable finite volume scheme, we can gain new insights and a more
satisfying explanation as to why the aforementioned approaches
work.\newline 

As it turns out, a pressure blip, exactly analogous to SPH, is
produced when using the Lagrange plus Eulerian remap version of the
PPM code, PPMLR~\cite{colella-woodward-ppm,vh1,cmhog}. Remap schemes
involve a Lagrangian advection step in which the cells move, followed
by a conservative remap onto the original Eulerian grid. We find
(agreeing with the argument of Davies et al.~\cite{davies93}) it
highly unlikely that two fundamentally different approaches result in
the same erroneous features. Interestingly, the Eulerian version of
the PPM (PPMDE) and indeed, other Eulerian schemes
\cite{toro-book,leveque-book} do not exhibit this anomaly. Lagrangian
finite difference codes have traditionally fallen out in favour of
their Eulerian counterparts and we are led to conjecture therefore
that the difference between the two versions of the PPM scheme can
provide an answer to origin of the pressure jump in SPH.\newline 

This work is outlined as follows. In Sec.~\ref{sec:evidence}, we use a
one-dimensional shock tube problem to provide numerical evidence to
the claim that PPMLR exhibits, qualitatively the same errors as the
SPH pressure jump at the contact discontinuity. In
Sec.~\ref{sec:blip-discussion}, we use the spatiotemporal behaviour of
the SPH error to draw an analogy with ``wall-heating'' errors for
traditional finite difference schemes and argue that the pressure jump
is a result of a spurious entropy generation in the initial transient
phase of shock formation. In Sec.~\ref{sec:explanation}, through a
comparison of PPMLR and PPMDE schemes, a lack of diffusion in the
material wave (contact) is highlighted as the source of the error and
we demonstrate how it may be eliminated for PPMLR by using a more
diffusive Lagrangian advection step. Finally, in
Sec.~\ref{sec:sph_application}, we use these ideas to propose a minor
modification to the GSPH scheme, where the requisite dissipation is
added through a suitable choice of an approximate Riemann solver
(similar to the method proposed by Shen et al.~\cite{ShenYan10}). The
scheme is applied to standard test problems in one and two dimensions
to demonstrate it's effectiveness in mitigating the pressure jump for
flows with strong shocks. We summarize this work in
Sec.~\ref{sec:summary}, with conclusions drawn from this work and
suggestions for possible extensions. All numerical results presented
in this manuscript are generated using the code (SPH2D)\footnote{SPH2D
  is available at https://bitbucket.org/kunalp/sph2d}, which is freely
available for validation and use. 

\section{Numerical evidence of the pressure blip}
\label{sec:evidence}
In this section, we provide numerical evidence for the existence of
the pressure discontinuity when using SPH and the Lagrange plus remap
version of the Piecewise Parabolic Method (PPMLR
\cite{colella-woodward-ppm,vh1}) finite difference code. The error is
particularly severe for SPH and generally arises across a density
gradient (contact discontinuity). We consider two one-dimensional
shock tube problems that work well to highlight the anomalous
behaviour. The initial conditions are defined with the the left (l)
and right (r) states displayed in Table.~\ref{tab:cases}
\begin{table}[h]
  \centering
    \begin{tabular}{|l | c | c | c| c| c| r|} \hline
      Test & $\rho_l$ & $p_l$ & $u_l$ & $\rho_r$ & $p_r$ &$u_r$\\ \hline 
      Sod's shock-tube & 1& 1& 0& 0.125& 0.1& 0 \\ \hline 
      Blast-wave & 1 & 1000 & 0 & 1 & 0.01 & 0 \\ \hline 
    \end{tabular}
  \caption{Initial data for the test problems to highlight the pressure
    discontinuity}
  \label{tab:cases}
\end{table}
The first test is the famous Sod's shock tube~\cite{sod1978-survey}
problem. This test represents the bare minimum a numerical scheme for
the compressible Euler equations should hope to resolve. The initial
conditions result in a right moving contact discontinuity sandwiched
between a left moving rarefaction and a right moving shock wave. The
solution is self similar with four constant states separated by the
three waves. The second test is a more stringent version, referred to
as the blast-wave problem, in which the magnitude of the initial
pressure jump ($p_l/p_r$) is $10^5$. The wave structure is similar
with a strong, right facing shock wave ($M = 198$) moving into the low
pressure gas. \newline For the SPH results, we use our SPH2D
\cite{Puri2013-I} code that implements the variational formulation
described by Price \cite{price2012}. Thermal conduction is turned off
for both test problems to highlight the errors at the contact
discontinuity. For the PPMLR method, we have used John Blondin's VH-1
\cite{vh1} and Jim Stone's CMHOG \cite{cmhog} codes independently to
verify our results. Both codes implement the PPMLR algorithm proposed
by Colella and Woodward \cite{colella-woodward-ppm}.
\subsection{Test $1$: Sod's shock-tube problem}
\label{sec:sod-shock-evidence}
\begin{figure}[!h]
  \begin{center}
    %\showthe\columnwidth
    \includegraphics[width=\columnwidth]{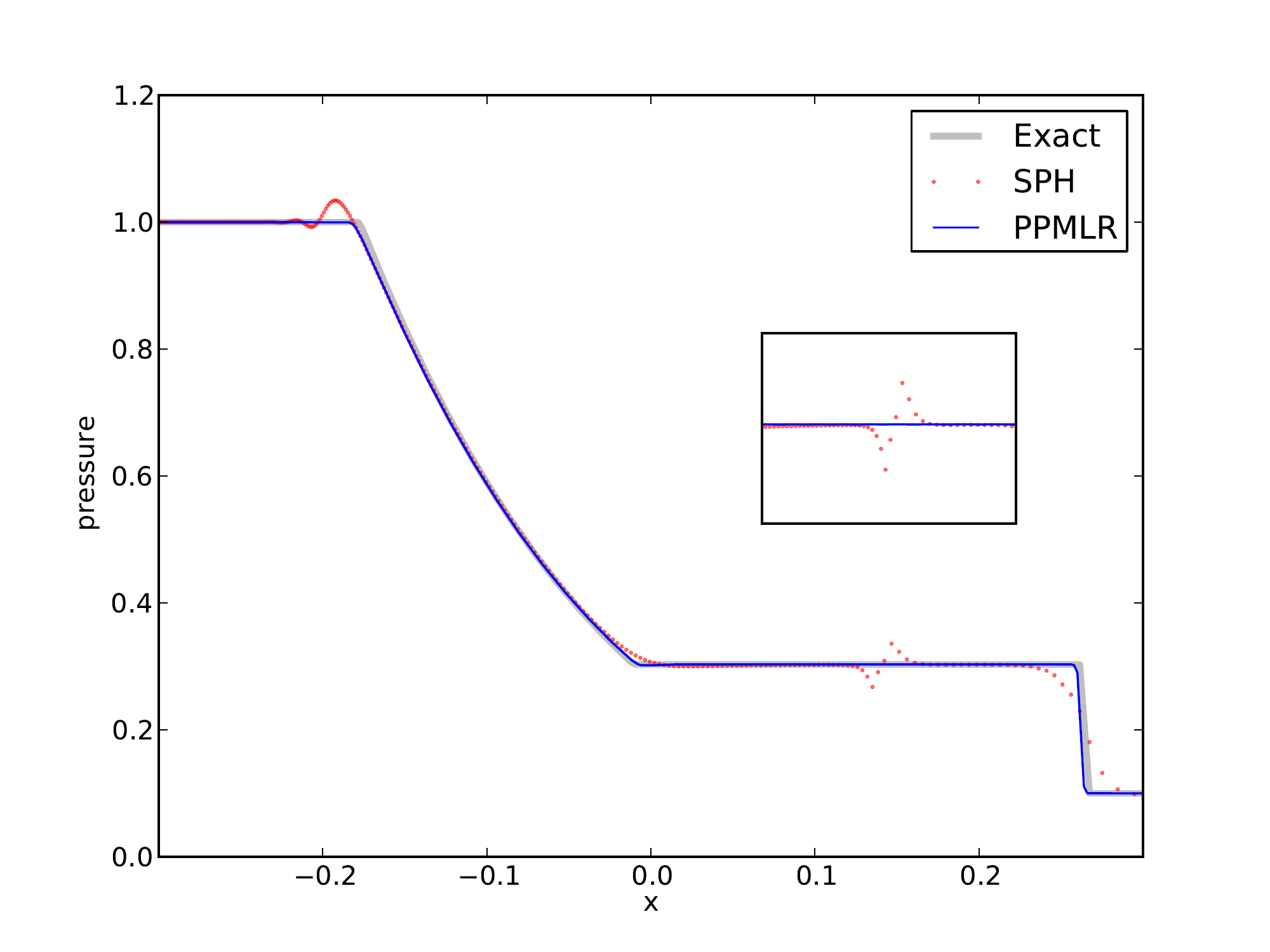}
    \caption{Numerical pressure for Sod's shock-tube problem using SPH
      (dots), PPMLR (solid blue line) at $t = 0.15s$, compared with
      the exact solution (solid black line). The pressure ``blip'' is
      clearly visible in the SPH solution at the contact $x\approx
      0.17$. A close-up of the solution around the contact is shown in
      the inset.}
    \label{fig:sodshock-pressure-evidence}
  \end{center}
\end{figure}
The numerical pressure profiles for the SPH and PPMLR schemes is shown
in Fig.~\ref{fig:sodshock-pressure-evidence}. The SPH simulation was
performed using a total of $450$ particles. Initially, $400$ particles
were placed to the left of the initial discontinuity ($x = 0$) with
spacing $\Delta x_l = 0.00125$. The remaining $50$ particles were
placed to the right of $x = 0$, with a spacing of $\Delta x_r =
0.01$. The particle mass was set equal to the inter-particle spacing
$\Delta x_l$ so that $m/\Delta x$ reproduces the desired
density. For the PPMLR results, we used a total of $500$ grid cells
(zones). The pressure discontinuity is clearly visible for the SPH
results. A close up of the solution in the vicinity of the contact is
shown in the inset. For this relatively simple problem, PPMLR does not
exhibit the anomalous pressure jump at the contact. The situation is
different for the blast-wave problem discussed next.
\subsection{Test $2$: Blast-wave problem}
\label{sec:blastwave-evidence}
\begin{figure}[!h]
  \begin{center}
    %\showthe\columnwidth
    \includegraphics[width=\columnwidth]{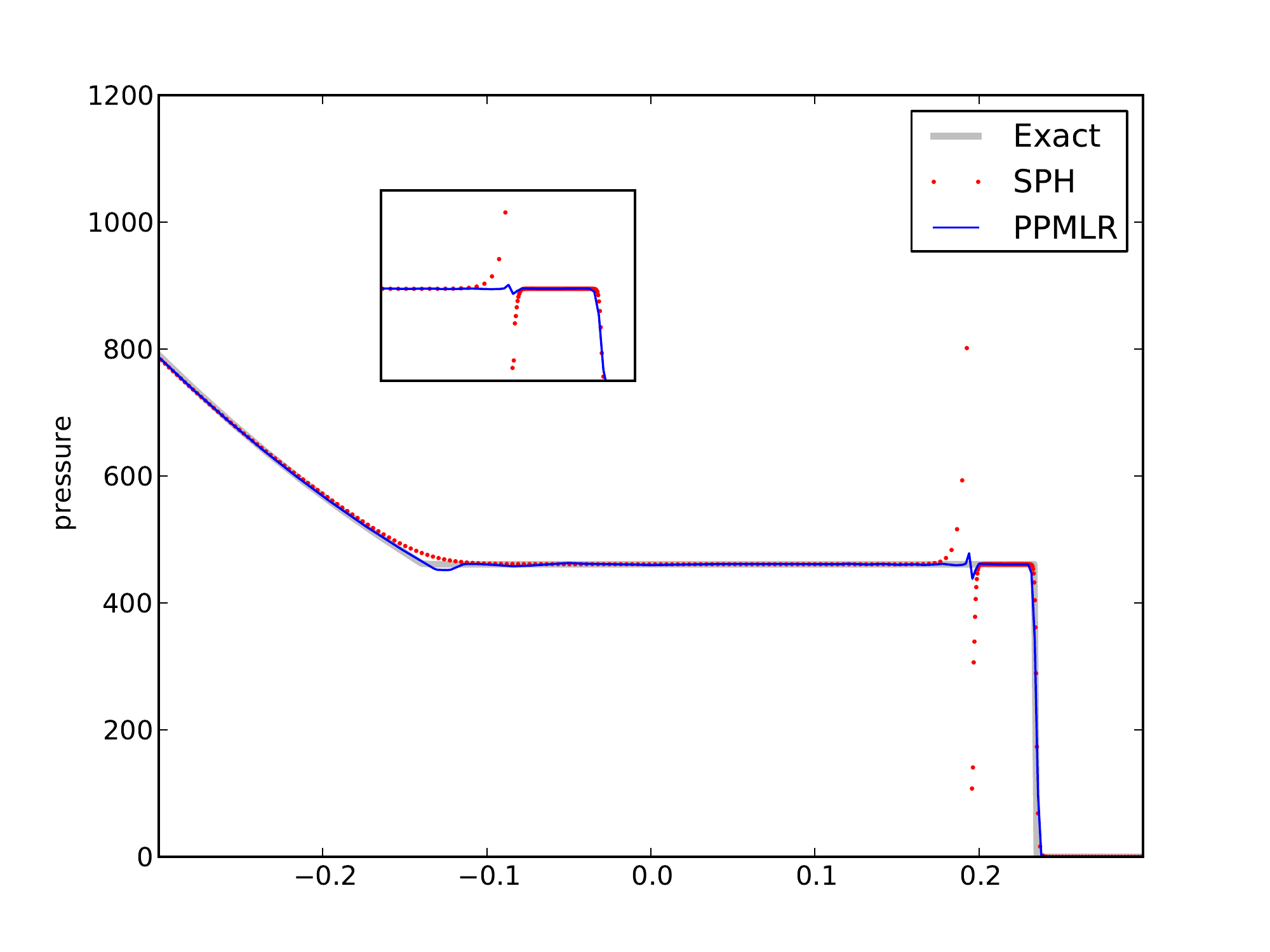}
    \caption{Numerical pressure for the blast-wave problem using SPH
      (dots), PPMLR (solid blue line) at $t = 0.01s$ compared with the
      exact solution (solid black line). The pressure blip is visible
      for both SPH and PPMLR (see inset) schemes.}
    \label{fig:blastwave-pressure-evidence}
  \end{center}
\end{figure}
Numerical pressure profiles for the SPH and PPMLR schemes are shown in
Fig.~\ref{fig:blastwave-pressure-evidence}. We notice that the blip
which was absent in the PPMLR solution for the Sod's shock-tube
problem is now present. Another striking feature is the huge jump in
the pressure for the SPH solution. The results were generated using a
total of $500$ particles for SPH and $500$ grid cells for PPMLR. At
the outset, it may seem that the SPH results are no-where in
comparison to PPMLR but this is not the case. This is because the
PPMLR scheme has an inherent diffusion for the thermal energy which
works to dissipate the error with time. Recall that we explicitly
switched off the thermal conduction for the SPH scheme. With a small
amount of thermal conduction, the results of the two schemes are
similar as
\begin{figure}[!h]
  \begin{center}
    %\showthe\columnwidth
    \includegraphics[width=\columnwidth]{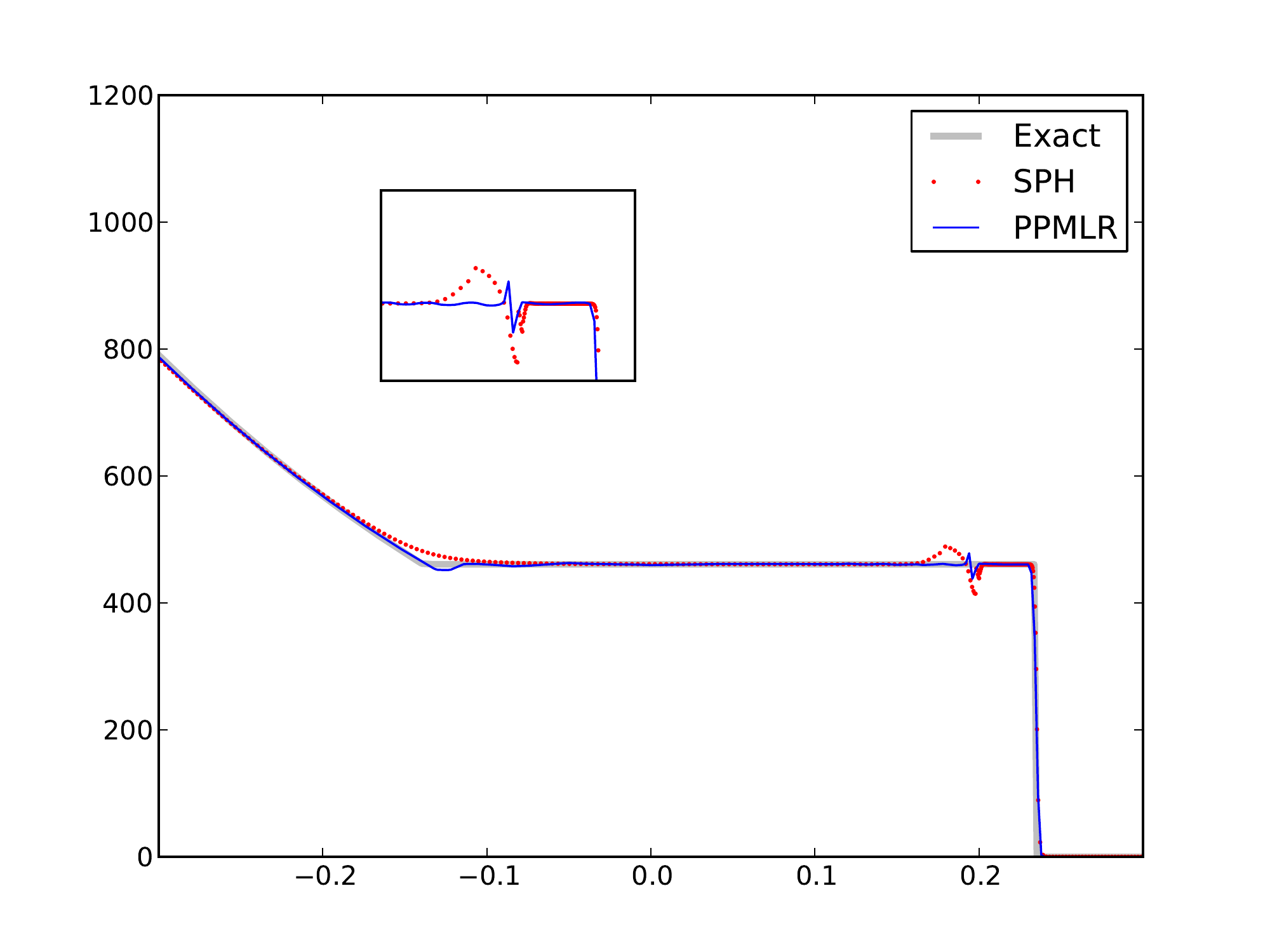}
    \caption{Numerical pressure for Blast-wave problem using SPH with
      thermal conduction (dots), PPMLR (solid blue line) at $t =
      0.01s$ compared with the exact solution (black line). The
      pressure blip produced by each of the schemes is now similar.}
    \label{fig:blastwave-pressure-evidence2}
  \end{center}
\end{figure}
can be seen in Fig.~\ref{fig:blastwave-pressure-evidence2}, where the
magnitude of the SPH pressure jump is dramatically reduced and is only
slightly larger than that of the PPMLR scheme. Having established the
presence of the numerical error for both schemes, we now proceed to a
discussion on the nature of the error and provide an explanation for
it's occurrence.

\section{A discussion on the error}
\label{sec:blip-discussion}
In the previous section, we provided numerical evidence to support our
original claim that the ubiquitous pressure jump at the contact
discontinuity for SPH solutions, occurs for a class of finite
difference schemes. In particular, the Lagrange plus remap version of
the Piecewise Parabolic Method (PMPLR) results in an error remarkably
similar to SPH. We note that although code comparisons for SPH and PPM
in different contexts have previously been conducted
(\cite{davies93,oshea-etal05,sphvgrids,taskeretal08}), this error has
rather strangely gone unnoticed or has not been reported. It is
possible that the comparisons were carried out with the fully Eulerian
version of PPM, as two step (Lagrange plus remap) codes have
traditionally fallen out in favour of their fully Eulerian
counterparts. Indeed, Woodward and Colella
\cite{woodward-colella1984}, in their comparison of numerical schemes
for flows with strong shocks showed that the cell-centred, direct
Eulerian version of the PPM scheme (PPMDE) was the most accurate. In a
more recent study, Pember and Anderson \cite{pember-anderson00} argue
otherwise, stating that the two approaches yield generally equivalent
results. Nevertheless, direct Eulerian schemes are undoubtedly more
prevalent and higher order versions of these schemes do not exhibit
the SPH-like pressure jump at the contact, as can be verified by any
of the schemes presented in the monographs by Toro \cite{toro-book}
and LeVeque \cite{leveque-book}. We believe the differences between
the remap and direct Eulerian finite difference schemes could provide
insight into the nature of the pressure jump in SPH. We would like to
remind the reader that although we know of a suitable ``fix'' in the
form of thermal conduction, we are looking for a consistent
explanation for it's origin and a justification for the myriad
approaches outlined in the introduction.

\subsection{The nature of the error}
\label{sec:error-nature}
\begin{figure}[!h]
  \begin{center}
    %\showthe\columnwidth
    \includegraphics[width=\columnwidth]{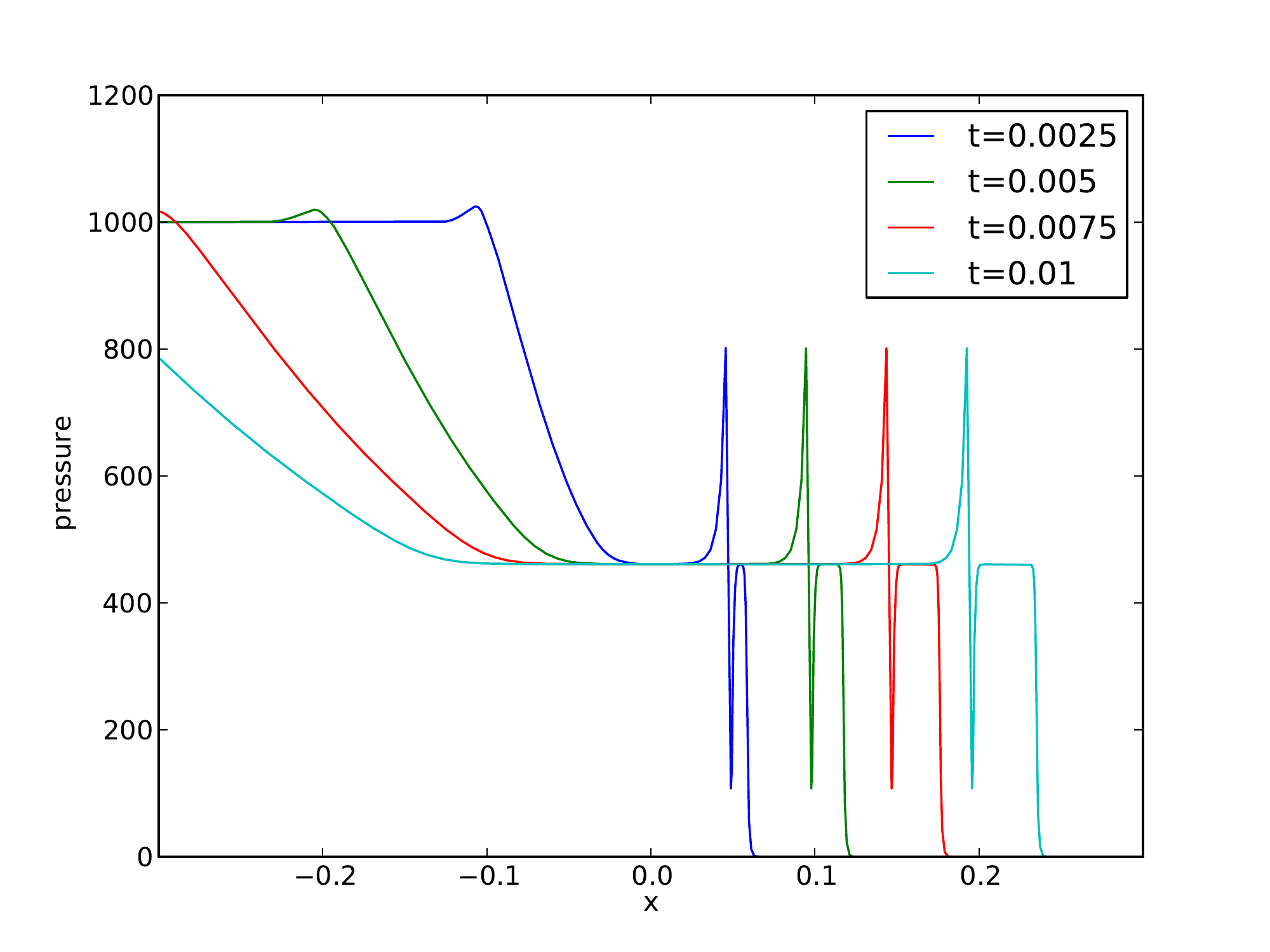}
    \caption{Numerical pressure for Blastwave problem at the time
      instants $t=0.0025$, $t = 0.005$, $t = 0.0075$ and $t=0.01$
      seconds, when simulated with SPH. Once the pressure ``blip'' is
      generated, it is advected without dissipation. Thermal
      conduction was explicitly turned off to highlight this
      behaviour.}
    \label{fig:blastwave-pressure-thist}
  \end{center}
\end{figure}

A natural question to ask of a numerical scheme is convergence to the
physically correct solution. The pressure jump at the contact, being
clearly erroneous, raises valid questions as to the behaviour of the
error with the spatial resolution. It is instructive therefore, to
catalogue known features of the pressure jump in the SPH context. We
continue with the strong shock problem of
Sec.~\ref{sec:blastwave-evidence} as the canonical example exposing
this behaviour for SPH. Thermal conduction is switched off to avoid
cosmetic smoothing of the results. Concerning the question of
numerical convergence, we first examine the error as we increase the
number of particles.
\begin{figure}[!h]
  \begin{center}
    %\showthe\columnwidth
    \includegraphics[width=\columnwidth]{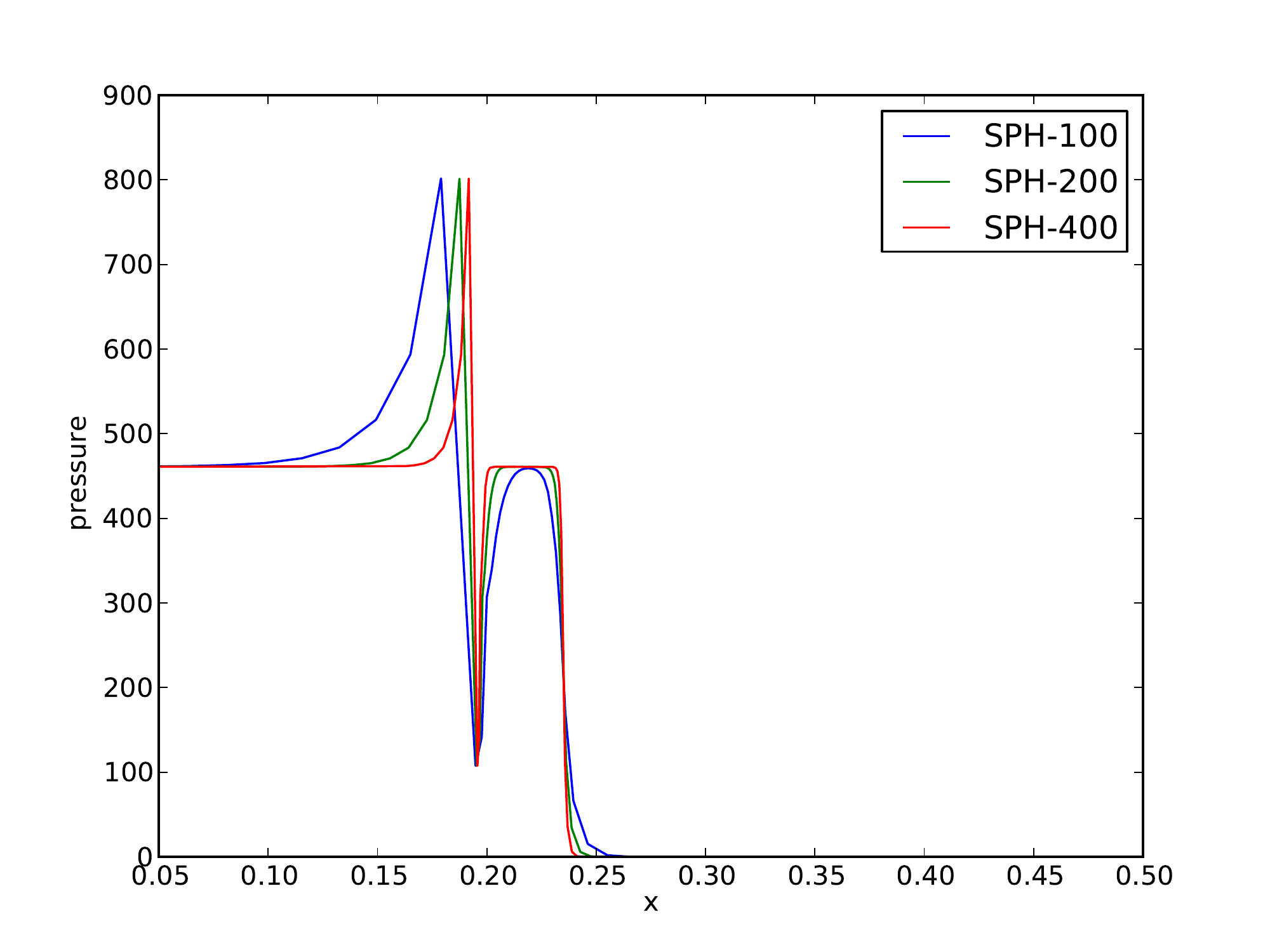}
    \caption{Numerical pressure for Blastwave problem using SPH with
      different resolutions. The solution profiles are expectedly more
      crisp for higher resolutions. The spread of the error reduces
      with increasing resolution but peak magnitude of the error
      remains constant.}
    \label{fig:blastwave-pressure-convergence}
  \end{center}
\end{figure}
Fig.~\ref{fig:blastwave-pressure-convergence} shows the results for
the blastwave problem when we have used $100$, $200$ and $400$
particles respectively. The shock transition region is sharper with
higher resolution as expected. We notice that the spread of the error
\emph{reduces} with increasing resolution but the peak point-wise
error remains \emph{constant}. Convergence in $L_{\infty}$ is
therefore not possible in SPH. Convergence in $L_{1}$ with a
convergence rate or approximately $2$ has been observed previously
\cite{springel2010}. The temporal behaviour of the error can be
studied through Fig.~\ref{fig:blastwave-pressure-thist}, which shows
snapshots of the pressure profile at the times $t=0.0025$, $t =
0.005$, $t = 0.0075$ and $t=0.01$. These results highlight another
feature of the error.
Namely, once the error is created, the pressure jump simply advects
with the flow and neither grows nor attenuates without explicit
thermal conduction. Artificial viscosity has no discernible effect on
the solution. This is not surprising when we consider that the
artificial viscosity is added to generate the correct entropy jump
across the shock and therefore, has no effect at the contact
discontinuity. To summarize, the pressure jump at the contact
discontinuity, in the context of SPH has the following properties:
\begin{itemize}
  \item The error, once created is simply advected with the material
    velocity.
  \item With the increase of spatial resolution, the spread of the
    error decreases whilst maintaining a constant peak magnitude.
  \item Artificial viscosity has no role to play in suppressing or
    mitigating it's effect.
\end{itemize}

\subsection{The relation to wall-heating errors}
\label{sec:wall-heating}
A multi-valued pressure in the presence of an accurate density profile
results, in a corresponding jump in the thermal energy, via the
equation of state. The pressure jump in SPH can therefore be thought
of as a spurious heating/cooling of the fluid. For the numerical
solution of the compressible Euler equations, ``wall-heating'' is the
canonical term given to errors that result in a spurious rise in
thermal energy (heating). The problem has generated considerable
interest and has received the attention of several
researchers~\cite{RiderWallHeating}. The term was coined by
Noh.~\cite{noh1978-wallshock} when he considered the problem of shock
reflection in planar, cylindrical and spherical geometries. The reason
for our interest in these errors is the similarity it bears with the
SPH errors at the contact discontinuity.  For instance, under mesh
refinement, Noh observed that the overheating decreases in spatial
range while maintaining it's peak point-wise value. This leads Noh to
conjecture that the error is built into the exact solution of the
modified viscous equations. He further argues that any numerical
method based on ``shock-smearing'' (artificial viscosity) will
demonstrate this error. By considering the asymptotic solution of the
governing equations with artificial viscosity,
Menikoff~\cite{Menikoff94} argues that entropy errors, resulting in
spurious heating/cooling are generated not only for shock reflection,
but also for shock interaction or when a shock passes through a change
in mesh spacing. The errors are generated over a short transient phase
as a result of the numerical width of the shock. After the initial
transient phase, the entropy errors are \emph{frozen} and simply
advected with the fluid and the only recourse is to add thermal
conduction which leads to a diffusion of the energy. A similar
analysis in Lagrangian coordinates was carried out by Shen et
al.~\cite{ShenYan10}. Once again, entropy errors were observed to
occur over a short, initial transient phase of shock
interaction/reflection.\newline These observations for wall heating
errors in the context of traditional finite volume methods bear a
striking similarity to the SPH pressure/entropy errors. Thus, we
conjecture that the SPH errors are a form of spurious heating that
occurs in an initial transient phase. Once the entropy errors are
generated, diffusion of entropy (thermal conduction) is the only
recourse to mitigate it's effect. 

\section{An Explanation for the Error}
\label{sec:explanation}
We argued that the pressure jump at the contact discontinuity for SPH
has symptoms of wall-heating errors previously observed for
traditional finite difference/volume codes. Essentially, an entropy
error is generated during the initial transient phase of shock
formation, the magnitude of which is independent of the spatial
resolution. After the initial transient, the error is convected along
the particle trajectories without dissipation. Monaghan and
Gingold.~\cite{monaghan-gingold-shock} had originally ascribed this
anomalous behaviour for SPH, to generic ``starting'' errors. More than
twenty years later, Tasker et al.~\cite{taskeretal08} had also
suggested that the discontinuous initial conditions give rise to an
entropy error. Since the errors are generated at start-up, and
subsequently passively advected with the particles, thermal conduction
is required to mitigate it's effect. This was the also the conclusion
drawn by Noh.~\cite{noh1978-wallshock} when he proposed an artificial
heat flux for finite difference schemes. While this reasoning serves
to justify the artificial conductivity approach in treating the error,
it sheds no light onto the origins of the error. Towards this aim, we
adopt a different perspective by studying grid-based schemes.\newline
Recall that in Sec.~\ref{sec:evidence}, the errors for the Lagrange
plus Eulerian remap version of the PPM scheme (PPMLR) was shown to be
qualitatively similar to those of SPH. Agreeing with the reasoning of
Davies et al.~\cite{davies93}, we find it highly unlikely that two
fundamentally different schemes (SPH and PPMLR) would result in the
same erroneous features. This leads us to believe that both schemes
solve a similar modified equation, the solution of which exhibits the
heating and corresponding pressure jump at the contact. A scaling
argument similar to Noh.~\cite{noh1978-wallshock} shows that the
magnitude of the error is independent of the spatial resolution. This
justifies to some extent the claim that the error is built into the
exact solution of the discrete SPH equations. It is therefore
reasonable to assume that a solution to the problem within the finite
volume context using PPMLR might provide clues for a similar
resolution in SPH. This can be done by comparing the Lagrangian plus
remap PPM scheme, PPMLR, with it's direct Eulerian counterpart, PPMDE,
for which the error is absent.
\subsection{PPMLR and PPMDE}
\label{sec:ppm}
The Piecewise Parabolic Method (PPM) \cite{colella-woodward-ppm} is a
high order, Godunov finite difference method that has, as it's
building block, a third order advection scheme. Along with the
ENO/WENO type schemes \cite{JiangShu96}, PPM is considered to be a
highly accurate method for the compressible Euler equations
\cite{woodward-colella1984}. The scheme, as originally proposed by
Colella and Woodward~\cite{colella-woodward-ppm}, can be formulated to
follow either Lagrangian or Eulerian hydrodynamics. Although the two
equation sets are mathematically the same, their numerical solutions
exhibit differences which we are interested in.\newline In what
follows, we focus on the development of the one-dimensional PPM scheme
since multi-dimensional extensions are constructed with the
dimensional splitting approach. Thus, essential details of the method
and the error producing mechanism in particular are contained within
the one-dimensional scheme. In a such a scheme, the conserved
variables are the specific volume $1/\rho$, velocity $u$, and the
specific energy $\hat{e} = e + \frac{1}{2} u^2$ for the Lagrangian
formulation, and density $\rho$, momentum $\rho u$, and total energy
$\rho \hat{e}$, for the Eulerian formulation. Both versions of the PPM
scheme (PPMLR, PPMDE) advance the solution (vector of conserved
variables $\boldsymbol{U}$) over physical zones or cells. The generic
$i^{\text{th}}$ cell has it's center at $x_i$, left and right faces at
$x_{i-\frac{1}{2}}$ and $x_{i+\frac{1}{2}}$ respectively and $\Delta
x_i = x_{i+\frac{1}{2}} - x_{i-\frac{1}{2}}$ denotes the cell volume.
\subsubsection{PPMLR}
\label{sec:ppmlr}
The conservative equations for Lagrangian hydrodynamics are given as
\begin{subequations}
  \label{eq:lagrangian-hydrodynamics}
  \begin{align}
  \tau_t - u_{\xi} &= 0 \\
  u_t + p_{\xi} &=0 \\
  \hat{e}_t + (pu)_{\xi} &=0
  \end{align}
\end{subequations}
where $\tau = 1/\rho$ is the specific volume, $\hat{e} =
\frac{1}{2}u^2 + e$ is the total energy per unit mass and the time
derivative is to be understood as a derivative moving with the fluid
(material derivative) $\frac{d}{dt}(*) = \frac{\partial}{\partial
  t}(*) + u \nabla(*)$. $\xi$ is the \emph{mass} coordinate, which is
related to the spatial coordinate $x$, through the transformation
$d\xi = \rho dx$. The system is hyperbolic with eigenvalues $\lambda_1
= -C$, $\lambda_2 = 0$ and $\lambda_3 = C$, where $C = (\gamma p
\rho)^{\frac{1}{2}}$ is the Lagrangian sound speed. The convective
terms are absent in this formulation which is reflected as a wave of
speed $0$.\newline In PPMLR, the procedure to advance the solution is
carried out in two steps. In the first step, piecewise parabolic
interpolations of the pressure, velocity and density are used to
compute the effective \emph{left} and \emph{right} states for a
Riemann problem between two adjacent cells. Since the zone edge is
moving with the fluid velocity, the input state is determined purely
by the acoustic modes $\lambda_1$ and $\lambda_3$. With the input
(\emph{left}, \emph{right}) states constructed from the parabolic
reconstructions, a Riemann problem is solved to calculate fluxes
through the cell boundaries. The vector of conserved variables are
updated using the conservative differencing equations:
%j+\frac{1}{2}
\begin{subequations}
  \label{eq:ppmlr-advection}
  \begin{align}
    x_{j+\frac{1}{2}}^{n + 1} &= x_{j+\frac{1}{2}}^{n} + \Delta t
    \overline{u}_{j+\frac{1}{2}} \\ \tau_{j+\frac{1}{2}}^{n
      + 1} &= \frac{1}{\Delta m_j}\left(x_{j+\frac{1}{2}}^{n + 1} -
    x_{j-\frac{1}{2}}^{n + 1}\right) \\ u_{j+\frac{1}{2}}^{n
      + 1} &= u_{j+\frac{1}{2}}^{n} + \frac{\Delta t}{\Delta
      m_j}\left(\overline{p}_{j-\frac{1}{2}} -
    \overline{p}_{j+\frac{1}{2}}\right)
    \\ \hat{e}_{j+\frac{1}{2}}^{n + 1} &= u_{j+\frac{1}{2}}^{n} +
    \frac{\Delta t}{\Delta
      m_j}\left(\overline{u}_{j-\frac{1}{2}}\overline{p}_{j-\frac{1}{2}}
    - \overline{u}_{j+\frac{1}{2}}\overline{p}_{j+\frac{1}{2}}\right)
  \end{align}
\end{subequations}
In these equations, $\overline{p}$ and $\overline{u}$ denote the
intermediate pressure and velocity that results from the solution to
Riemann problem at a zone edge. Convection is introduced through the
cell motion. After the advection step, the solution is conservatively
re-mapped onto the original Eulerian grid. A necessary condition for
high-order Godunov schemes, non-linearity is introduced by using
limiters and monotonicity constraints in the piecewise parabolic data
reconstruction.
\subsubsection{PPMDE}
\label{sec:ppmde}
The one-dimensional equations for Eulerian, inviscid hydrodynamics are
given as
\begin{subequations}
  \label{eq:eulerian-hydrodynamics}
  \begin{align}
  \rho_t + (\rho u)_x &= 0 \\
  (\rho u)_t + (\rho u^2 + p)_x &= 0\\
  (\rho \hat{e})_t + (\rho \hat{e} u + pu)_x &=0
  \end{align}
\end{subequations}
where the symbols have the same meaning and $(*)_t$ is the partial
derivative with respect to time. The system of equations is again
hyperbolic with eigenvalues $\lambda_1 = u-c$, $\lambda_2 = u$ and
$\lambda_3 = u + c$, where $c = \rho C$, is the Eulerian sound
speed. The procedure to update the solution in the one-step, direct
Eulerian formulation is essentially the same as PPMLR. Piecewise
parabolic interpolations of the dependent variables are used to
compute effective \emph{left} and \emph{right} states for Riemann
problems between adjacent cells. Zone fluxes, computed from the
Riemann solution are used in a conservative differencing scheme.
\begin{subequations}
  \label{eq:ppmde-step}
  \begin{align}
    \boldsymbol{U}_{j+\frac{1}{2}}^{n+1} &=
    \boldsymbol{U}_{j+\frac{1}{2}}^{n} + \frac{\Delta t}{\Delta x_j}
    \left( \boldsymbol{F}(\overline{\boldsymbol{U}}_{j-\frac{1}{2}}) -
    \boldsymbol{F}(\overline{\boldsymbol{U}}_{j+\frac{1}{2}}) \right)
    \\ \overline{\boldsymbol{U}}_{j+\frac{1}{2}} &\approx
    \frac{1}{\Delta
      t}\int_{t^n}^{t^{n+1}}\boldsymbol{U}(x_{j+\frac{1}{2}}, t) dt
  \end{align}
\end{subequations}
The difference is in the construction of the input states for the
Riemann problem. This is now more complicated than the Lagrangian case
as there may now be as many as $3$ and as few as $0$ waves impinging
on a zone edge from a given side. A consequence of this is that in
general, for the input state at a given zone edge, contributions from
each wave family must be accounted for \cite{colella-woodward-ppm}.

\subsection{Diffusion in the material wave}
\label{sec:ppmde-diffusion}
Both PPMLR and PPMDE employ piecewise parabolic interpolations of the
cell centered density, pressure and velocity to construct the input
left and right states as integral averages over the characteristic
domain of dependence. The domain of dependence for a given wave family
is defined by tracing back the path of the wave if it impinges on the
zone edge from a given side. For the Lagrangian formulation, we have
two waves corresponding to the acoustic modes, $\lambda_1$ and
$\lambda_3$. The material wave is absent as the cells are assumed to
move with the local fluid velocity ($\lambda_2 = 0$). For the Eulerian
formulation, waves from each of the three families can impinge on an
edge from a given side. The input state in this case is constructed
such that the amount of wave associated with each family of
characteristics transported across a zone edge is correct up to terms
of second order \cite{colella-woodward-ppm}. Thus, the additional
material wave must be accounted for in the Eulerian formulation.
Diffusion across this wave is the main difference between the two
versions of PPM. Indeed, Eulerian Godunov schemes are known to be more
diffusive at the contact when compared to Lagrangian
formulations. This was highlighted by Woodward and Colella
\cite{woodward-colella1984} and later by Pember and
Anderson~\cite{pember-anderson00}, when they compared remap and direct
Eulerian finite volume schemes.
\begin{figure}[!h]
  \begin{center}
    %\showthe\columnwidth
    \includegraphics[width=\columnwidth]{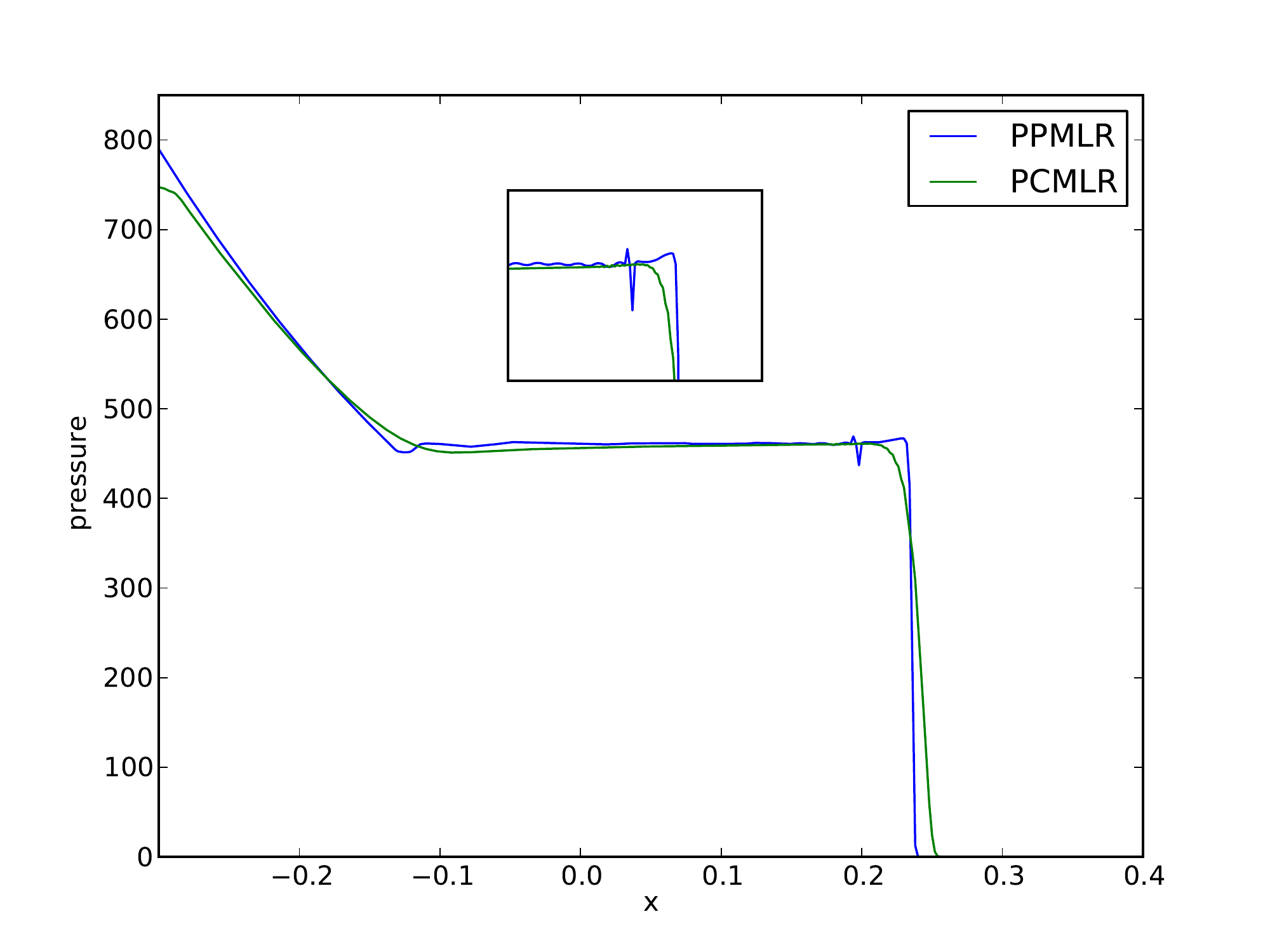}
    \caption{Results for the blast-wave problem using a remap code
      with piecewise parabolic (PPMLR) and piecewise constant (PCMLR)
      interpolations. The entropy error (pressure jump) is eliminated
      with when using the diffusive PCMLR method.}
    \label{fig:bwave-ppmlr-pcmlr}
  \end{center}
\end{figure}
This suggests a \emph{lack} of dissipation is actually the cause of
the SPH-like entropy error for PPMLR as observed in
Sec.~\ref{sec:evidence}. It was suggested by Jim Stone (private
communication, August 2013) that the low dissipation in PPM causes
these ``start-up'' errors. The discontinuous initial conditions gives
rise to additional waves on a discrete level which is captured by the
low dissipative schemes like PPMLR. One is then tempted to verify the
hypothesis by constructing a more diffusive version of the PPMLR
scheme. The easiest way to do this is to use a piecewise constant
reconstruction instead of the parabolic reconstruction used in PPMLR.
Results for the blast-wave problem using such a scheme (PCMLR) is shown
in Fig.~\ref{fig:bwave-ppmlr-pcmlr}. The solution is expectedly less
crisp than PPMLR but remarkably, the pressure jump at the contact is
eliminated. We note that it is not possible for the remap phase of
PPMLR to introduce the error. This is because remapping can be viewed
as a projection and is inherently a diffusive process. As a result, no
new extrema can occur in this step. The error is therefore generated
in the Lagrangian advection phase.\newline What does dissipation in
the material wave look like?  To answer this question, we consider the
eigenstructure of the equations in the Lagrangian formulation
(Eq.~\ref{eq:lagrangian-hydrodynamics}). The right and left
eigenvectors for this hyperbolic system is given as (\cite{Rider94})
\begin{equation}
  \label{eq:lagrangian-eigenstructure}
    R = 
    \begin{pmatrix}
      1 & 1 &1 \\ C &0 &-C \\ uC -p &\frac{p}{\gamma -1} &-uC -p
    \end{pmatrix},
    \,\,\,R^{-1} = 
    \begin{pmatrix}
      \frac{1}{2\gamma}& \frac{1}{2C} + \frac{u(\gamma-1)}{2\gamma p} & \frac{1-\gamma}{2\gamma p} \\
      \frac{\gamma-1}{\gamma}& \frac{u(\gamma-1)}{\gamma p} & \frac{\gamma-1}{\gamma p} \\
      \frac{1}{2\gamma}& -\frac{1}{2C} + \frac{u(\gamma-1)}{2\gamma p} & \frac{1-\gamma}{2\gamma p}
    \end{pmatrix}
\end{equation}
where $C = (\gamma p \rho)^{\frac{1}{2}}$ is the Lagrangian sound
speed. A conservative finite volume scheme with a general diffusive
flux contribution can be defined as
\begin{equation}
  \label{eq:lagrangian-diffusive-flux}
  \left(\boldsymbol{F}_{j+\frac{1}{2}}\right)_{\text{diss}} = -
  \frac{1}{2}\sum_{k=1}^{3}\hat{\boldsymbol{r}}_k
  |\hat{\lambda}_k|\left(\hat{\boldsymbol{r}}_k^{-1} \cdot \Delta
  \hat{\boldsymbol{U}}_{j+\frac{1}{2}} \right),
\end{equation}
where the caret denotes a suitably averaged value at the zone
interface $x_{j+\frac{1}{2}}$. This is the numerical flux for
\emph{linearized} schemes such as Roe's scheme~\cite{Roe81} and is
applicable to SPH~\cite{monaghan1997-riemann,purigsph13}. The
diffusive flux in Eq.~\ref{eq:lagrangian-diffusive-flux} computes
\emph{jumps} across each wave family. The magnitude of the jump (wave
strength), $\alpha = \hat{\boldsymbol{r}}_k^{-1} \cdot \Delta
\hat{\boldsymbol{U}}$, is weighted by the wave speed. The final
contribution to the conserved variables is determined by the right
eigenvector for that wave family. For the Lagrangian scheme, this
contribution vanishes for the material wave ($k = 2$), since
$\lambda_2 = 0$. This is the contribution we are interested in, the
algebraic form of which is given by
\begin{equation}
  \label{eq:lagrangian-material-wave-dissipation}
  \left(f_{j+\frac{1}{2}}^2\right)_{\text{diss}} =
  -|\lambda_2|\frac{1}{2}\,\begin{bmatrix} \frac{1}{2C}
  +\frac{u(\gamma-1)}{2\gamma p}& \frac{u(1-\gamma)}{\gamma p} &
  -\frac{1}{2C} +\frac{u(\gamma-1)}{2\gamma p}\end{bmatrix}
  \cdot \begin{bmatrix} \Delta \hat{\tau} \\ \Delta\hat{u} \\ \Delta
    \hat{e} \end{bmatrix} \begin{bmatrix} 1 \\ 0 \\ \frac{p}{\gamma
      -1} \end{bmatrix}
\end{equation}
Due to the structure of the right eigenvector ($\boldsymbol{r}_2$ in
Eq.~\ref{eq:lagrangian-eigenstructure}), the dissipation acts on the
first and third components of $\boldsymbol{U}$. For the Lagrangian
formulation, these are the specific volume $\tau = 1/\rho$, and total
energy per uni mass $\hat{e} = \frac{1}{2}u^2 + e$ respectively. Thus,
dissipation in the material wave would result in an additional density
and energy diffusion \emph{simultaneously} that is absent in a purely
Lagrangian formulation.  

\section{Application to SPH}
\label{sec:sph_application}
The lack of dissipation in the material wave, coupled with the low
diffusion of PPMLR is responsible for the entropy errors, and hence
the SPH-like pressure jump at the contact. The introduction of
dissipation helped eliminate the error for the finite volume remap
code in Sec.~\ref{sec:explanation}. It is then reasonable to assume
that an improvement of results can be expected for SPH if this
dissipation is somehow introduced. Indeed, this has been the adopted
practice within the SPH community, with dissipation often introduced
directly through thermal conduction
\cite{sigalotti2006-shock,price2008-khi,Wadsley08,Garcia-Senz01012009,rosswog09-review,valcke10,read12},
or via surrogate means such as using a smoother estimate to define
particle volume~\cite{saitoh12,hopkins12}, relaxing initial conditions
\cite{valcke10,read10} and a modification to the equation of state
\cite{lanzafame09}. The problem with the dissipation introduced in
these schemes is that they appear serendipitous and their reasoning
belies the simplicity of the SPH formulation. The requirement that
dissipation should act across the material wave provides a consistent
explanation as to why the aforementioned approaches work. From
Eq.~\ref{eq:lagrangian-material-wave-dissipation}, we know that a
combination of dissipation in the density and energy variables is
required to suppress the entropy errors. Dissipation for velocity
(artificial viscosity) has no role to play. This was verified
numerically in Sec.~\ref{sec:blip-discussion}. Armed with this
knowledge, we can attempt to introduce the requisite dissipation in a
consistent manner for SPH, thereby validating our hypothesis.
\subsection{Adding diffusion to SPH}
\label{sec:sph_diffusion}
%% Having established SPH as a scheme with very low dissipation, it is
%% understandable that the scheme is particularly susceptible to
%% generating large entropy errors that result in the ubiquitous
%% pressure blip. We argued that dissipation, in particular, that
%% which results from an omission of the material wave in the
%% Lagrangian formulation needs to be introduced. The dissipation must
%% be added in the density and energy equations and we propose a way
%% to do this.\newline 

We consider the GSPH formulation \cite{inutsuka-riemann} with an
approximate Riemann solver. We have shown (\cite{purigsph13}) that
with a suitable choice of an approximate Riemann solver, this
formulation is equivalent to a variational SPH scheme with artificial
dissipation and thermal conduction. The advantage of this formulation
is the explicit control of the dissipation through numerical fluxes
akin to finite difference/volume schemes. The discrete SPH equations
in this formulation, for the density, velocity and thermal energy are
given as
\begin{subequations}
\label{eq:gsph-equations-simple}
\begin{align}
  \label{eq:gsph-summation-density}
  \rho_a &= \sum_{b\in\mathcal{N}(a)} m_b W_{ab}(h_a)\\
  \label{eq:gsph-simple-mom}
  \ddot{\boldsymbol{x}}_a &= -\sum_{b \in \mathcal{N}(a)}m_b
  p^*_{ab}\left(\frac{1}{\rho_a^2}\nabla W_{ab}(h_a) +
  \frac{1}{\rho_b^2}\nabla W_{ab}(h_b)\right) \\
  \label{eq:gsph-simple-enr}
  \dot{e}_a &= -\sum_{b \in \mathcal{N}(a)}m_b
  p^*_{ab}[\boldsymbol{v}_{ab}^{*} - \dot{\boldsymbol{x}}_a] \left(
  \frac{1}{\rho_a^2}\nabla W_{ab}(h_a) +
  \frac{1}{\rho_b^2}\nabla W_{ab}(h_b)\right)
\end{align}
\end{subequations}
where, the starred quantities ($p^*, \, \boldsymbol{v}^*$) are the
intermediate state arising from the solution of a Riemann problem
between two interacting particles. The solution to the Riemann problem
introduces the minimum and necessary dissipation required to stabilize
the scheme.\newline We follow the approach proposed by Shen et
al.~\cite{ShenYan10} by constructing a hybrid scheme in which a
regular Riemann solver is used in the momentum equation equation, and
a \emph{diffusive} Riemann solver is used for the energy
equation. Recall that the flux vector for the conservative equations
in the Lagrangian formulation are $\left(-u, p, pu\right)$. Thus,
using a more \emph{dissipative} intermediate velocity is akin to
introducing dissipation in the density and energy equations. In
\cite{purigsph13}, we evaluated $5$ different approximate Riemann
solvers in Lagrangian coordinates for use with GSPH. From an analysis
of an accuracy test for the Euler equations, we find that the Harten,
Lax, van Leer and Einfeldt (HLLE)~\cite{Rider94} solver is a suitable
choice for the \emph{diffusive} approximate Riemann solver. For the
regular Riemann solver, we can use any one from the exact
\cite{vanleer-muscl5}, Ducowicz \cite{Ducowicz85} or Roe \cite{Roe81}
approximate Riemann solvers.
\begin{figure}[!h]
  \begin{center}
    %\showthe\columnwidth
    \includegraphics[width=\columnwidth]{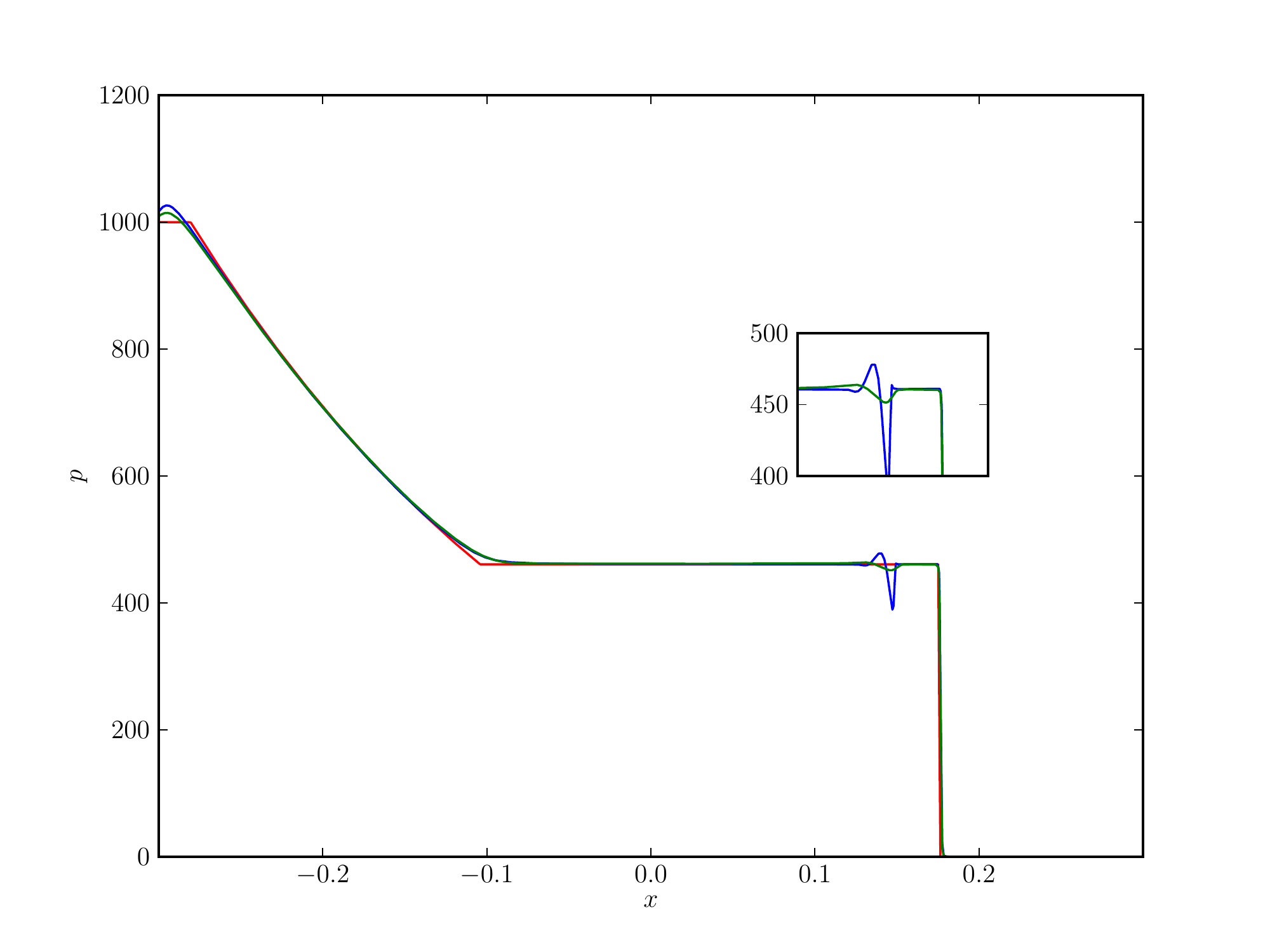}
    \caption{Numerical pressure profiles for the blastwave problem
      using standard GSPH (blue) and the hybrid GSPH (green) using the
      HLLE approximate Riemann solver, compared with the exact
      solution (ref). The HLLE solver is successful in suppressing the
      entropy errors as can be seen in the inset plot.}
    \label{fig:gsph-hybrid-blastwave-pressure}
  \end{center}
\end{figure}
We construct such a scheme, where the two approximate Riemann solvers
used are the van Leer exact and the HLLE approximate Riemann
solver. In particular, the diffusive contribution is constructed as
\begin{subequations}
  \label{eq:gsph-hybrid-vstar}
  \begin{align}
    u^* &= \frac{1}{t_f}(tu^*_{\text{regular}} +
    (t_f - t)u^*_{\text{diff}}), \\
    p^* &= \frac{1}{t_f}(tp^*_{\text{regular}} +
    (t_f - t)p^*_{\text{diff}}),
  \end{align}
\end{subequations}
where, $t_f$ is the final time in the simulation. This corresponds to
a linear blending of the two estimates with a more diffusive estimate
used in the initial stages of the computation. Since the errors are
expected to be generated at start-up, the blending avoids excessive
dissipation that may ruin the
solution. Fig.~\ref{fig:gsph-hybrid-blastwave-pressure} shows the
numerical pressure profiles for the standard (blue) and hybrid GSPH
(green), compared with the exact solution (red), when using $1000$
equal mass particles. As expected, the dissipation helps to suppress
the entropy error, with only a slight kink visible in the inset plot.
\begin{figure}[!h]
  \begin{center}
    %\showthe\columnwidth
    \includegraphics[width=\columnwidth]{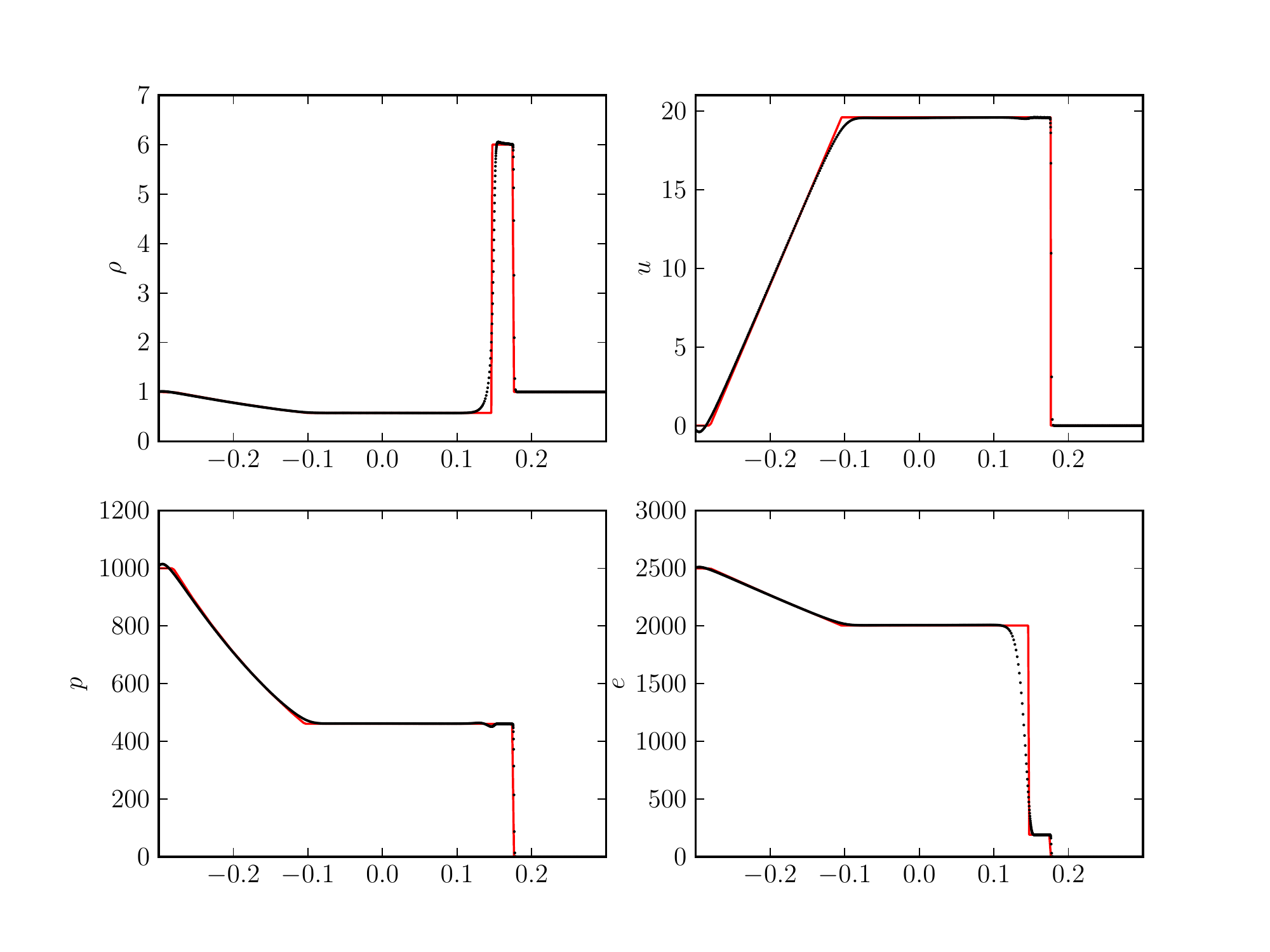}
    \caption{Numerical solution (dots) for the hybrid GSPH compared
      with the exact solution (red). The hybrid scheme shows good
      agreement with the exact solution with a negligible pressure
      jump. The dissipation has the desired effect of acting on the
      contact discontinuity.}
    \label{fig:gsph-hybrid-blastwave-all-variables}
  \end{center}
\end{figure}
The suppression of the pressure jump at the contact discontinuity does
not have an adverse effect on the profiles of the other physical
variables as can be seen in
Fig.~\ref{fig:gsph-hybrid-blastwave-all-variables}, which shows the
numerical solution for the hybrid GSPH scheme (dots), compared with
the exact solution (red line). The dissipation has the desired effect
of acting on the contact discontinuity as can be observed by the
slightly smeared density and thermal energy profiles.  near $x \approx
0.15$.

\begin{figure}[!h]
  \begin{center}
    %\showthe\columnwidth
    \includegraphics[width=\columnwidth]{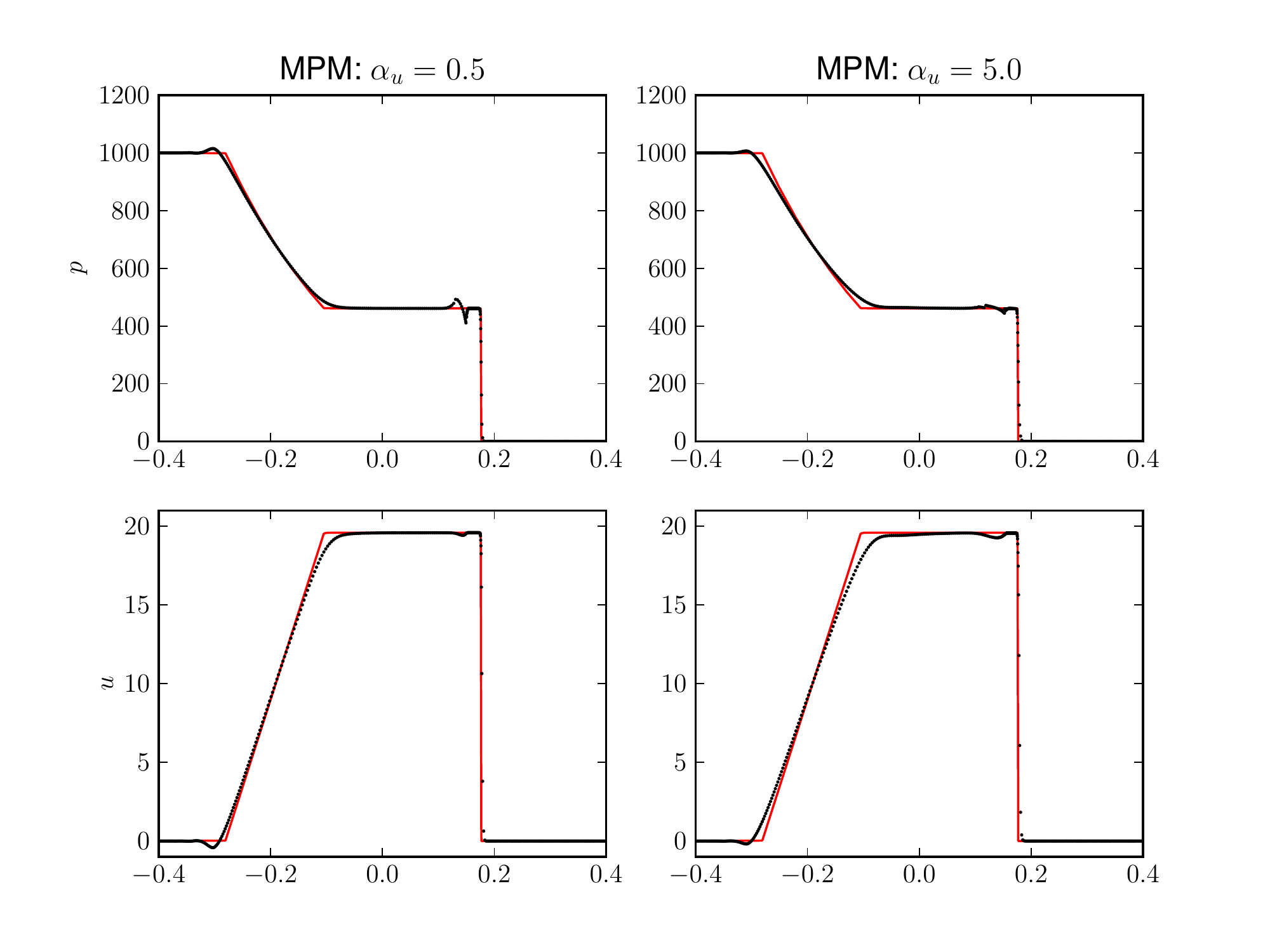}
    \caption{Effect of increasing the thermal conduction parameter
      $\alpha_u$ for the MPM scheme. Larger values of the parameter
      (right) mitigate the jump in the pressure (upper panel) but also
      generates a corresponding dip in the velocity profile (lower
      panel).}
    \label{fig:mpm-conduction-parameter}
  \end{center}
\end{figure}

We would like to point out that this method produces improved results
than with using a traditional scheme with larger thermal conduction
parameters. Fig.~\ref{fig:mpm-conduction-parameter} shows the result
of simply increasing the thermal conduction parameter $\alpha_u$, for
the variational scheme of Monaghan, Price and
Morris~\cite{monaghan1997-riemann,morris-monaghan1997-switch,price2012}.
Larger values of the thermal conduction parameter works to supress the
presure discontinuity as expected. However, results in an unwanted dip
in the velocity profile at the contact discontinuity. This behaviour
has been recently reported by Sirotkin and Yoh~\cite{sirotkin_yoh2013}
in their SPH scheme with approximate Riemann solvers. In comparison,
the hybrid GSPH scheme (cf
Fig.~\ref{fig:gsph-hybrid-blastwave-all-variables}) does not produce
this behaviour.

\subsection{Consequences of adding dissipation}
\label{sec:consequences}
The linear blending of the two estimates for $\boldsymbol{v}^*$
through Eq.~\ref{eq:gsph-hybrid-vstar} was shown to work well to
suppress the pressure jump for the one-dimensional blast-wave
problem. Since the errors are expected to be generated at start-up,
it's use can be detrimental for long time simulations.
\begin{figure}[!h]
  \begin{center}
    %\showthe\columnwidth
    \includegraphics[width=\columnwidth]{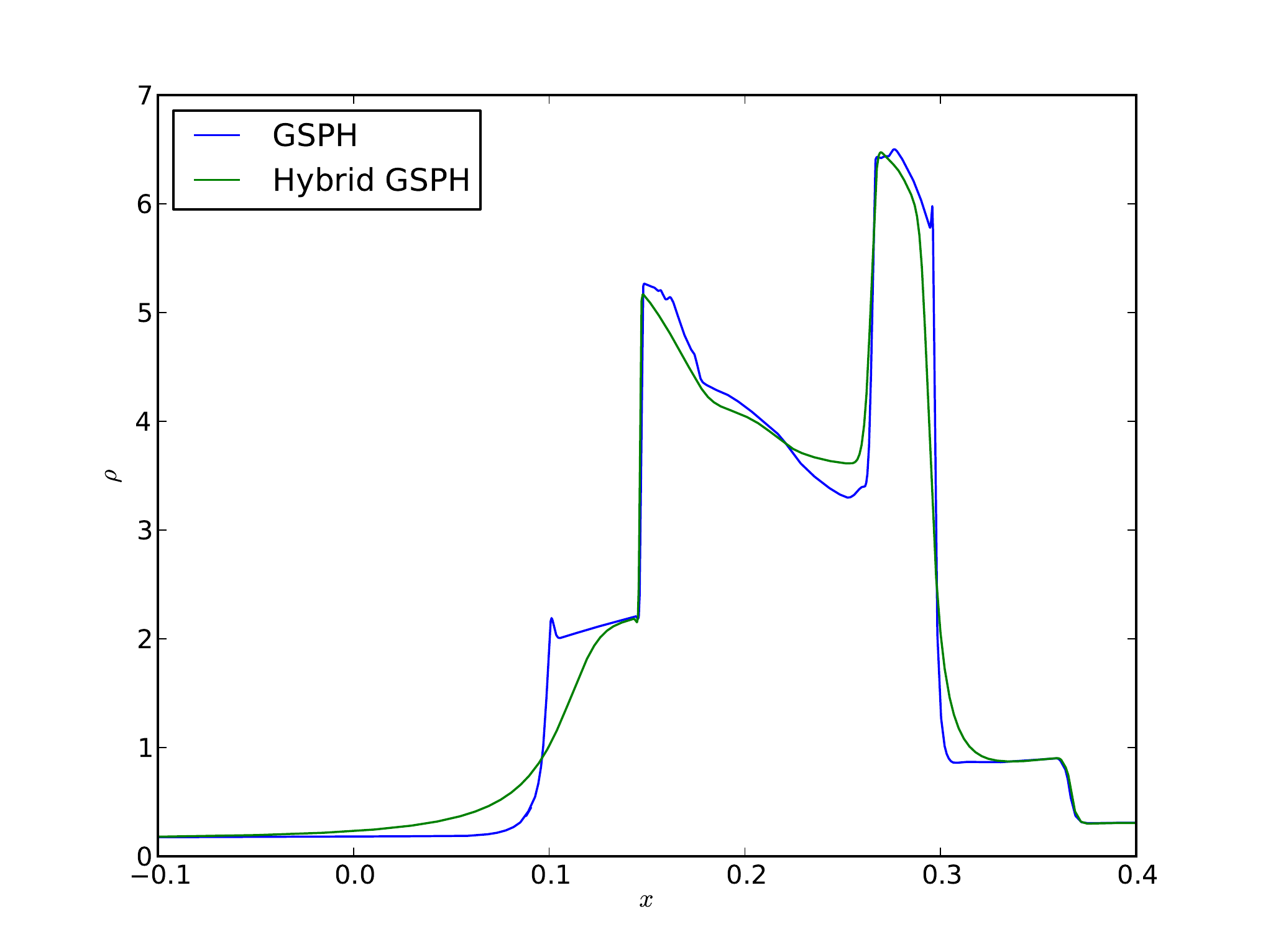}
    \caption{Density profiles at $t = 0.038$ for the Woodward and
      Colella blast-wave problem using standard GSPH (blue) and the
      hybrid modification (green) using
      Eq.~\ref{eq:gsph-hybrid-vstar}. The contact discontinuities near
      $x = 0.1$ and $x = 0.3$ are heavily smeared for the hybrid
      scheme and the solution within $x \in [0.15, 0.3]$ has lost some
      detail.}
    \label{fig:wcblast-linear-blending}
  \end{center}
\end{figure}
For example, Fig.~\ref{fig:wcblast-linear-blending} shows the density
profiles for the Woodward and Colella blast-wave problem
\cite{woodward-colella1984,Puri2013-I} using standard GSPH (blue) and
the hybrid modification (green) at the time $t = 0.038$. The extra
dissipation in the hybrid scheme has resulted in an excessive smearing
of the contact discontinuities near $x = 0.1$ and $x = 0.3$, and a
loss of detail within the region $x \in [0.15, 0.3]$. Note that adding
dissipation to the material wave (contact discontinuity) is exactly
what we set out to do, although Eq.~\ref{eq:gsph-hybrid-vstar} results
in an over diffusive scheme. This can be corrected by defining the intermediate states as
\begin{subequations}
  \label{eq:gsph-hybrid-vstar-exp}
  \begin{align}
  u^* &= u^*_{\text{regular}} +
  e^{-\alpha\frac{t}{t_f}}\left(u^*_{\text{diff}} -
  u^*_{\text{regular}}\right) \\
  p^* &= p^*_{\text{regular}} +
  e^{-\alpha\frac{t}{t_f}}\left(p^*_{\text{diff}} -
  p^*_{\text{regular}}\right)
  \end{align}
\end{subequations}
This corresponds to an exponential decay with time, for the diffusive
component. The \emph{parameter} $\alpha$, controls the rate of decay
(growth) of the two velocity estimates, with higher values resulting
in a more rapid decay (growth).
%% \begin{figure}[!h]
%%   \begin{center}
%%     %\showthe\columnwidth
%%     \includegraphics[width=\columnwidth]{exponential-blending.pdf}
%%     \caption{Blending function for $\boldsymbol{v}^*_{\text{DUCO}}$
%%       and $\boldsymbol{v}^*_{\text{HLLE}}$ given by
%%       Eq. \ref{eq:gsph-hybrid-vstar-exp}. The parameter $\alpha$
%%       controls the rate of decay for the dissipative estimate
%%       $\boldsymbol{v}^*_{\text{HLLE}}$}
%%     \label{fig:exponential-blending}
%%   \end{center}
%% \end{figure}
\begin{figure}[!h]
  \begin{center}
    %\showthe\columnwidth
    \includegraphics[width=\columnwidth]{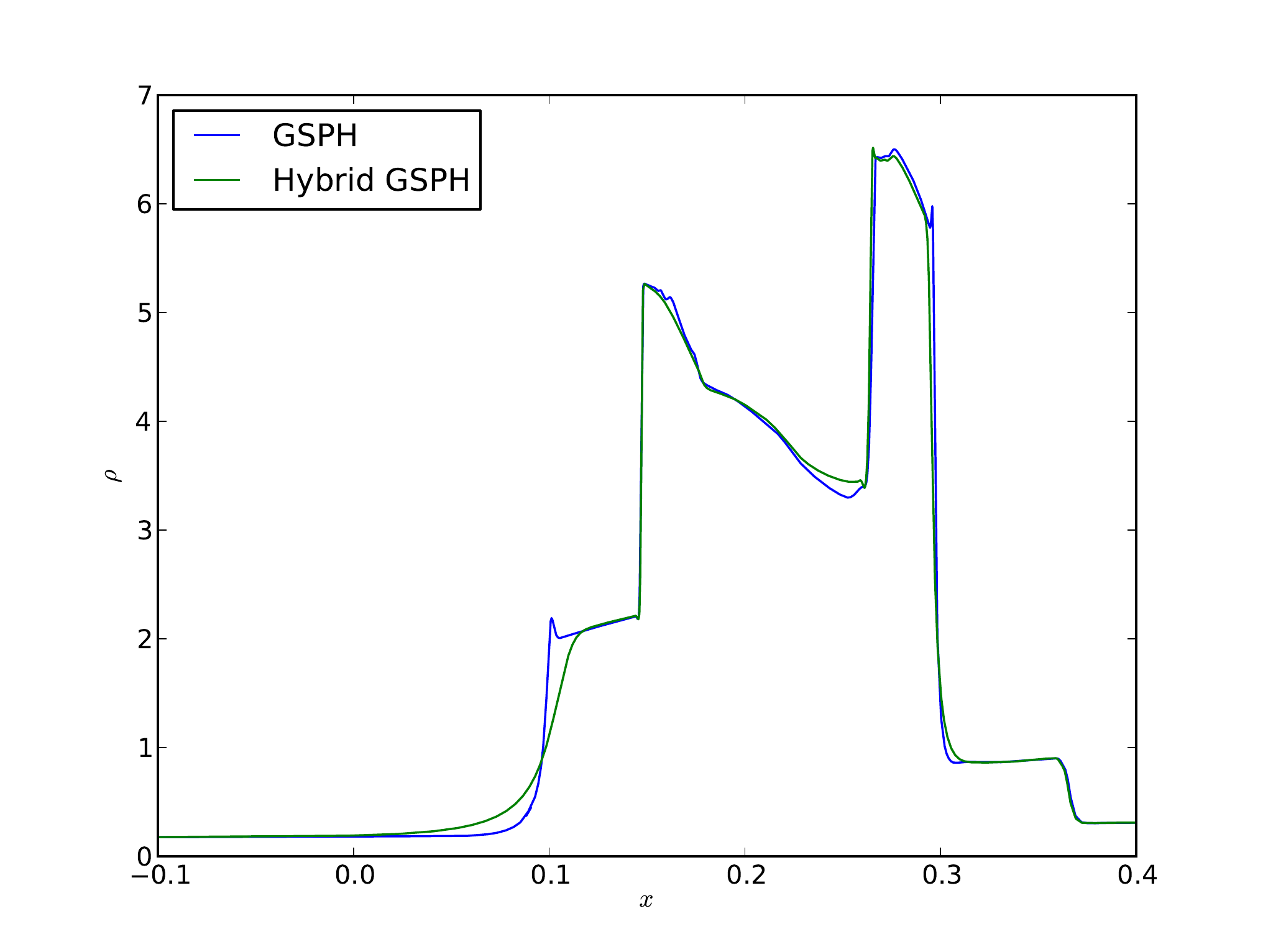}
    \caption{Density profiles at $t = 0.038$ for the Woodward and
      Colella blast-wave problem using standard GSPH (blue) and the
      hybrid modification (green) using
      Eq.~\ref{eq:gsph-hybrid-vstar-exp} and $\alpha = 10$. The
      solution is expectedly more crisp with a good agreement in the
      region $x \in [0.15, 0.3]$}
    \label{fig:wcblast-exp-blending}
  \end{center}
\end{figure}
Fig.~\ref{fig:wcblast-exp-blending} shows the density profiles at $t =
0.038$ for the same problem when we have this blending function with
$\alpha = 10$. The density profile for the hybrid GSPH (green) scheme
is expectedly more crisp with an improved agreement with the second
order GSPH (blue) scheme in the region $x \in [-0.15, 0.3]$. \newline

Fig.~\ref{fig:wcblast-exp-blending} shows the density profiles at $t =
0.038$ for the same problem when we have this blending function with
$\alpha = 10$. The density profile for the hybrid GSPH (green) scheme
is expectedly more crisp with an improved agreement with the second
order GSPH (blue) scheme in the region $x \in [-0.15, 0.3]$. \newline
We concede that a \emph{tuning} parameter to control dissipation
perhaps goes against the ethos of a GSPH scheme that is inherently
parameter free. While one can argue that the choice of the Riemann
solver itself in the GSPH scheme can be thought of as a
\emph{parameter}, the issue we want to highlight is the subtle role of
dissipation, that is needed for stability but criticized when used in
excess. An ideal scheme should use just the right amount of
dissipation for \emph{all} problems, the requisite amount, in turn,
ideally determined by the scheme itself (adaptive schemes). High-order
Godunov methods (MUSCL, PPM, ENO/WENO) are generally accepted to fit
this ideal. However, Quirk \cite{quirk94} famously pointed out several
instances where these schemes fail or produce erroneous
results. Moreover, he suggests that most of these errors can be
overcome by a judicious use of artificial dissipation. The trick is to
avoid a proliferation of \emph{tuning} parameters to determine the
requisite dissipation. We are faced with a similar conundrum while
constructing our hybrid GSPH scheme. Dissipation must be somehow
introduced into the purely inviscid equations and in this work we have
argued in favour of a specific form, acting across the energy and
density variables to suppress entropy errors. Given the transient
nature of the error, we are forced to introduce a \emph{parameter}
that limits the extra dissipation to when it is needed.

\subsection{Extension to higher dimensions}
\label{sec:2d}
\begin{figure}[!h]
  \begin{center}
    %\showthe\columnwidth
    \includegraphics[width=\columnwidth]{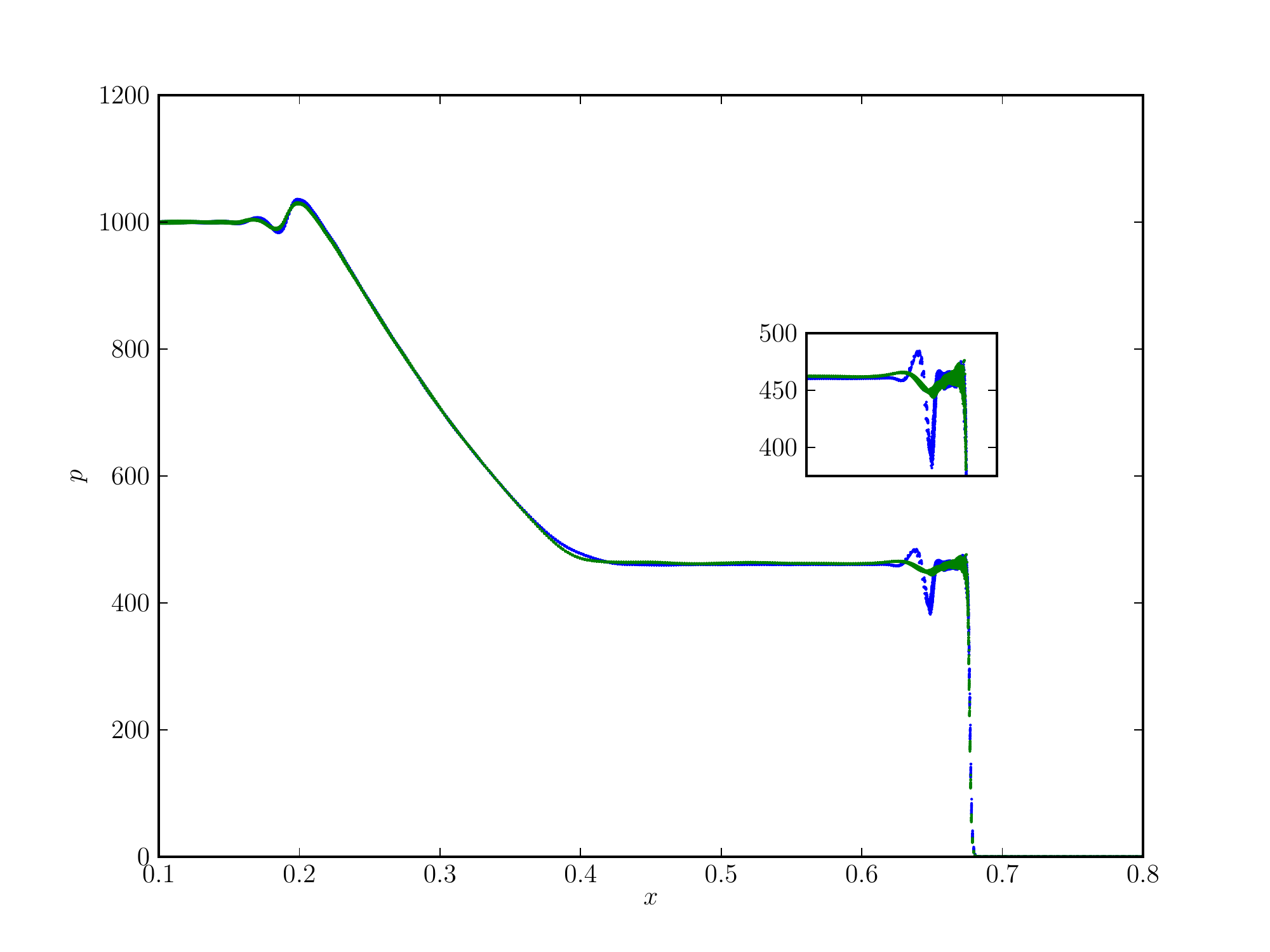}
    \caption{Numerical pressure for the hybrid GSPH (green) scheme
      with $\alpha = 3.0$ and with the choice of HLLE as the diffusive
      Riemann solver, compared with the standard second order GSPH
      scheme using the van Leer exact Riemann solver (blue). The
      results are exactly analogous to the one-dimensional case
      (Fig.~\ref{fig:gsph-hybrid-blastwave-pressure}), with the hybrid
      scheme working well to limit the pressure jump.}
    \label{fig:gsph-hybrid-blastwave2d-pressure}
  \end{center}
\end{figure}

\begin{figure}[!h]
  \begin{center}
    %\showthe\columnwidth
    \includegraphics[width=\columnwidth]{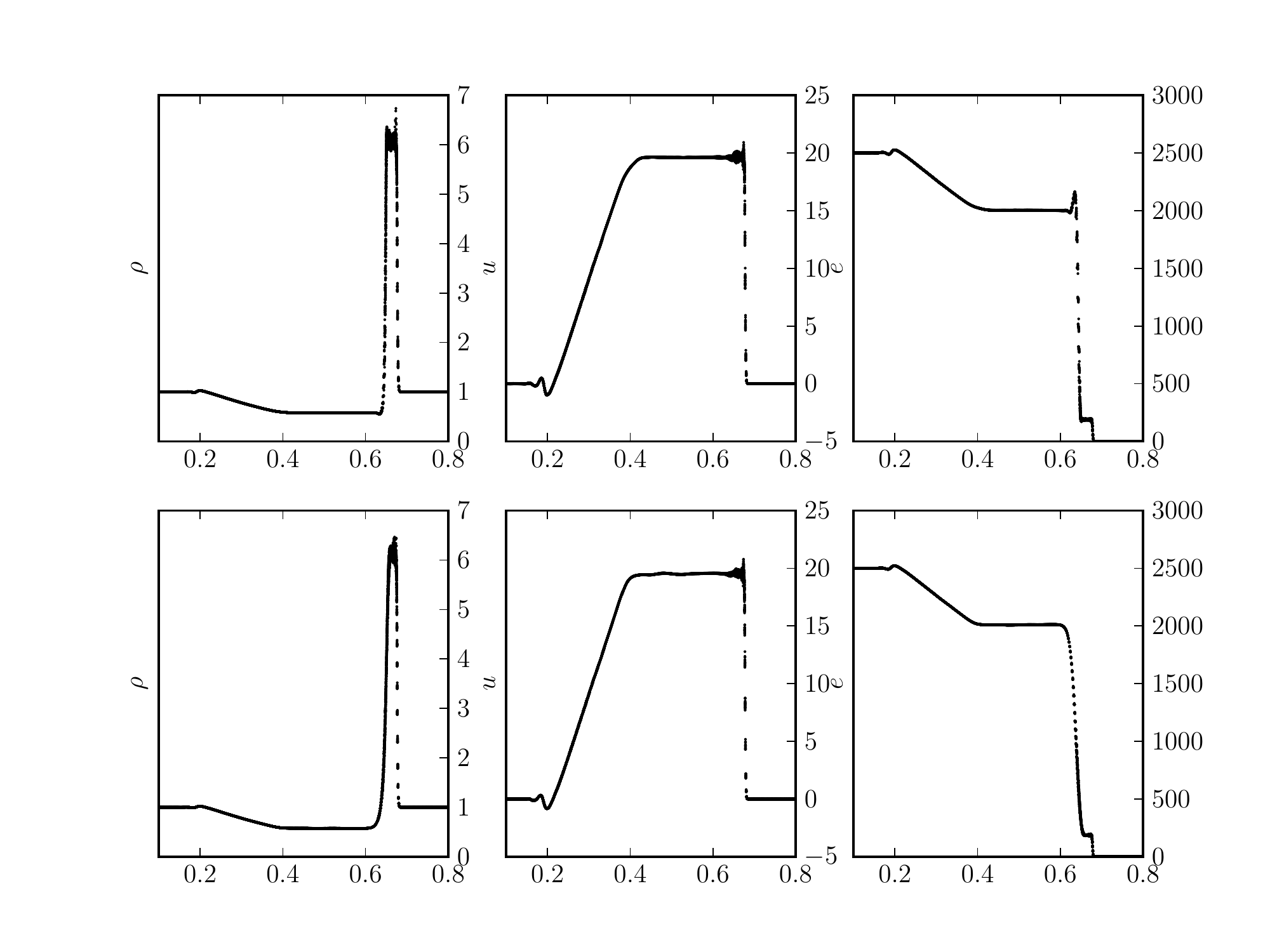}
    \caption{Numerical solution for the hybrid GSPH scheme (lower
      panel) with $\alpha = 3$ and with the choice of HLLE as the
      diffusive Riemann solver, compared with the standard second
      order GSPH scheme (upper panel). The hybrid scheme works to
      suppress the pressure blip
      (Fig.~\ref{fig:gsph-hybrid-blastwave2d-pressure}) and has no
      discernible adverse effect on the other physical variables. In
      particular, the spike in the thermal energy behind the contact
      is eliminated. Additionally, the contact discontinuity is
      slightly smeared for the density.}
    \label{fig:gsph-hybrid-blastwave2d-all-variables}
  \end{center}
\end{figure}

We used the one-dimensional blast-wave problem as the canonical test
highlighting the SPH entropy errors. The manner in which we chose to
introduce the dissipation is however, not limited to the
one-dimensional case. This can bee seen in
Fig.~\ref{fig:gsph-hybrid-blastwave2d-pressure}, which shows the
numerical (dots) pressure for the hybrid scheme (green) compared with
the reference second order GSPH scheme (blue) with the van Leer
Riemann solver, for a two-dimensional blastwave problem.
For the hybrid scheme, we use the exponential smoothing given by
Eqs.~\ref{eq:gsph-hybrid-vstar-exp} with $\alpha = 3.0$, and choose
the HLLE solver as the diffusive Riemann solver. The suppression of
the pressure blip has no discernible adverse effect on the other
variables as can be seen in
Fig.~\ref{fig:gsph-hybrid-blastwave2d-all-variables}, which shows the
density (left), velocity (center) and thermal energy (right) for the
hybrid GSPH scheme (lower panel), compared with a second order GSPH
scheme using the van Leer exact Riemann solver (upper panel).
The hybrid scheme eliminates the spike in the thermal energy behind
the contact. Additionally, analogous to the one-dimensional case, the
contact discontinuity is slightly smeared as can be seen in the
density plot around $x \approx 0.6$. The results were generated using
a total of $20000$ equal mass particles initially distributed in a
hexagonal close paced arrangement.

\section{Summary and further work}
\label{sec:summary}
In this work, we attempted to provide an explanation for the origin of
the ubiquitous pressure jump in SPH simulations of the compressible
Euler equations. The anomalous behaviour has been observed since the
dawn of SPH~\cite{monaghan-gingold-shock} and has received attention
ever since. It has been highlighted as a drawback of the method when
compared with traditional Eulerian
schemes~\cite{sphvgrids,taskeretal08}, and has led some researchers to
develop particle tessellation techniques as an alternative to
SPH~\cite{springel-moving-mesh10,hess-tessellation10}. Within the SPH
community, the use of dissipation, introduced via various means has
been the general recourse to mitigate the effects of the
error.\newline

Through an analogy with ``wall heating'' errors for finite
difference/volume codes, we argued that the pressure jump is a result
of entropy errors generated over an initial transient phase and
thereafter, passively advected with the particles. We highlighted that
a qualitatively similar error is present for the Lagrange plus remap
version of the Piecewise Parabolic Method (PPM) finite volume code
PPMLR. Through a comparison of PPMLR with it's direct Eulerian
counterpart, PPMDE, a lack of diffusion across the material wave was
identified as the origin for the error. By examining the
eigenstructure of the Lagrangian equations of motion, we showed that
the requisite diffusion needs to act on the density and energy
equations simultaneously. This explanation also justifies the myriad
techniques employed by different researchers to ``cure'' the
problem. Using our hypothesis, we construct a hybrid GSPH scheme that
introduces the requisite dissipation by using a more diffusive flux
for the energy equation. We verified our hypothesis by using the
blast-wave problem as the canonical test highlighting the pressure
anomaly in SPH. The results using the new scheme are shown to be
better than simply increasing the magnitude of thermal conduction for
SPH schemes that rely on explicit dissipation. A \emph{tuning}
parameter is introduced to limit the dissipation to the initial stages
of the computation.\newline

We expect the added dissipation to be disadvantageous for certain
problems. An example is the Sj\"ogreen's strong rarefaction
test~(1-2-3 problem \cite{toro-book}), for which, numerical
dissipation should be kept to a minimum. Indeed, one of the advantages
of the SPH artificial viscosity is the ability to switch it off
entirely when not required. Additionally, we believe that the hybrid
GSPH scheme can certainly be improved upon by using an alternative
hybridization to Eqs.~\ref{eq:gsph-hybrid-vstar-exp}. These equations
were constructed for the specific example to validate our hypothesis
concerning the origin of the spurious pressure jump in SPH. The
construction of an adaptive hybridization and validation for a general
suite of multi-dimensional problems is left as an area for future
investigation.

%\input{summary}
%\input{formulation}

%% References
%%
%% Following citation commands can be used in the body text:
%% Usage of \cite is as follows:
%%   \cite{key}          ==>>  [#]
%%   \cite[chap. 2]{key} ==>>  [#, chap. 2]
%%   \citet{key}         ==>>  Author [#]

%% References with bibTeX database:

\bibliographystyle{model6-num-names}
\newpage
\bibliography{entropy_errors}

%% Authors are advised to submit their bibtex database files. They are
%% requested to list a bibtex style file in the manuscript if they do
%% not want to use model1a-num-names.bst.

%% References without bibTeX database:

% \begin{thebibliography}{00}

%% \bibitem must have the following form:
%%   \bibitem{key}...
%%

% \bibitem{}

% \end{thebibliography}

\end{document}